\documentclass[a4paper,fleqn,usenatbib]{mnras}
\usepackage{newtxtext,newtxmath}
\usepackage[T1]{fontenc}
\usepackage{ae,aecompl}

\usepackage{graphicx}
\usepackage{amsmath}
\usepackage{amssymb}
\usepackage{float}
\usepackage{subfigure}
\usepackage{lipsum}
\usepackage{gensymb}
\usepackage{mhchem}
\usepackage{graphicx}
\usepackage{longtable}
\usepackage{rotate}
\usepackage{rotating}
\usepackage{mathtools}
\usepackage{pdflscape}
\usepackage{scalerel}
\usepackage{times}
\usepackage{verbatim}
\usepackage{tablefootnote}
\newif\ifAMStwofonts

\def\nustar{{\it NuSTAR}}
\def\chandra{{\it Chandra}}
\def\suzaku{{\it Suzaku}}
\def\athena{{\it Athena}}
\def\xrism{{\it XRISM}}
\def\asca{{\it ASCA}}
\def\sax{{\it BeppoSAX}}
\def\rxte{{\it RXTE}}
\def\swift{{\it Swift}}
\def\xmm{{\it XMM-Newton}}

\def\et{{et al.\ }}

\def\redchi{{\chi^2_\nu}}

\def\feka{{Fe~K$\alpha$}}
\def\feii{{Fe~\textsc{ii}}}
\def\fexxv{{Fe~\textsc{xxv}}}
\def\fexxvi{{Fe~\textsc{xxvi}}}

\def\arcs{{\hbox{$^{\prime\prime}$}}}

\def\cm{{\rm\thinspace cm}}
\def\erg{{\rm\thinspace erg}}
\def\eV{{\rm\thinspace eV}}

\def\kev{{\rm\thinspace keV}}

\def\Msun{\hbox{$\rm\thinspace M_{\odot}$}}
\def\ks{{\rm\thinspace ks}}
\def\s{{\rm\thinspace s}}

\def\ergpscmps{\hbox{$\erg\cm^{-2}\s^{-1}\,$}}

\title[Multi-epoch X-ray analysis of Mrk 478]{Multi-epoch X-ray spectral analysis of the narrow-line Seyfert 1 galaxy Mrk 478}

\author[S. G. H. Waddell et al.]{
S. G. H. Waddell,$^{1}$\thanks{E-mail: swaddell@ap.smu.ca}
L. C. Gallo,$^{1}$
A. G. Gonzalez,$^{1}$
S. Tripathi$^{1}$
and A. Zoghbi$^{2}$
\\
$^{1}$Department of Astronomy \& Physics, Saint Mary's University, 923 Robie Street, Halifax, Nova Scotia, B3H 3C3, Canada \\
$^{2}$Department of Astronomy, University of Michigan, 1085 South University Avenue, Ann Arbor, MI 48109-1107, US\\
}

\date{Accepted XXX. Received YYY; in original form ZZZ}

\pubyear{2019}

\begin{document}
\label{firstpage}
\pagerange{\pageref{firstpage}--\pageref{lastpage}}
\maketitle

\begin{abstract}
A multi-epoch X-ray spectral and variability analysis is conducted for the narrow-line Seyfert 1 (NLS1) active galactic nucleus (AGN) Mrk 478. All available X-ray data from \xmm\ and \suzaku\ satellites, spanning from 2001 to 2017, are modelled with a variety of physical models including partial covering, soft-Comptonisation, and blurred reflection, to explain the observed spectral shape and variability over the 16 years. All models are a similar statistical fit to the data sets, though the analysis of the variability between data sets favours the blurred reflection model. In particular, the variability can be attributed to changes in flux of the primary coronal emission. Different reflection models fit the data equally well, but differ in interpretation. The use of {\sc reflionx} predicts a low disc ionisation and power law dominated spectrum, while {\sc relxill} predicts a highly ionised and blurred reflection dominated spectrum. A power law dominated spectrum might be more consistent with the normal X-ray-to-UV spectral shape ($\alpha_{\rm ox}$). Both blurred reflection models suggest a rapidly spinning black hole seen at a low inclination angle, and both require a sub-solar ($\sim0.5$) abundance of iron. All physical models require a narrow emission feature at $6.7\kev$ likely attributable to \fexxv\ emission, while no evidence for a narrow $6.4\kev$ line from neutral iron is detected.
\end{abstract}


%%%%%%%%%%%%%%%%%%%%%%%%%%%%%%%%%%%%%%%%%%%%%%%%%%%%%%%%%%%%%%%%%%%%%%%%%%%%%%%%%%%%%%%%%%%%%%%%%%%%%%%%%%%%%%%%%%%%%%%%%%%%%%%%%%%%%%%%%%%%%%%%%%%%%%%%%%%%%%%%%%%%%%%%%%%%%%%%%%%%%%%%%%%%%%%%%%%%%%%%%%%%%%%%%%%%%%%%%%%%%%%%%%%%%%%%%%%%%%%%%%%%%%%%%%%%%%%%%%%%%%%%%%%%%%%%%%%%%%%%%%%%%%%%%%%%%%%%%

\section{Introduction}
\label{sect:intro}

Active galactic nuclei (AGN) are supermassive black holes that are actively accreting material. These objects are responsible for some of the most energetic phenomena in the Universe, and are typically highly variable across all wavelengths, from radio to $\gamma$-ray emission. The spectrum of these AGN will depend on the viewing angle through the obscuring torus (\citealt{Antonucci+1993, Urry+1995}). Seyfert 1 galaxies allow direct viewing of the central engine, whereas Seyfert 2 galaxies are viewed through the dusty torus.

Narrow-line Seyfert 1 (NLS1) galaxies are a sub-classification of Seyfert 1s that exhibit narrow Balmer lines and strong \feii\ emission lines in the optical region (\citealt{Osterbrock+1985, Goodrich+1989}). They are thought to contain lower mass black holes which are accreting material at a high rate, close to the Eddington limit (e.g. \citealt{Mathur2000, Gallo+2018}). In the X-ray regime, where the hottest photons located in the innermost regions are emitted, spectra for different objects have distinct similarities. Above $\sim2\kev$, the spectrum is usually dominated by a power law, produced by Compton up-scattering of UV accretion disc photons within a hot corona. Other spectral features include a soft excess below energies of $2\kev$ of disputed origin, and broadened emission lines attributed to the \feka\ emission line at $6.4\kev$ (e.g. \citealt{Fabian+1989}).

Despite the common appearance in AGN spectra, the physical origin of the X-ray soft excess is uncertain, and a variety of mechanisms may be responsible. A variety of models have been suggested, including partial covering (e.g. \citealt{Tanaka+2004}), soft-Comptonisation (e.g. \citealt{Done+2012}) and blurred reflection (e.g. \citealt{RossFabian+2005}). In the partial covering explanation, the soft excess could be an artifact of some obscuring material along the line-of-sight that absorbs some part of the intrinsic emission produced by the corona. As photons pass through this material, they are absorbed, producing deep absorption edges. Some of these photons are trapped by the Auger effect, but others are released, producing emission lines. The strengths of these lines are governed by the fluorescent yield of each element and the geometry of the absorbers around the primary emitter. Spectral variability can be explained by changes in the column density, covering fraction and ionisation of the absorbers without invoking changes in the intrinsic emission. The partial covering scenario has been used to explain the spectrum and variability found in low mass X-ray binaries (e.g. \citealt{Brandt+1996,Tanaka+2003}) and AGN (e.g. \citealt{Tanaka+2004,Miyakawa+2012,Gallo+2015}). This model typically includes a strong \feka\ absorption edge at $\sim7\kev$, a weak \feka\ line at $6.4\kev$, and absorption at low energies. A variety of column densities, ionisation states (from neutral to highly ionised) structures, (e.g. opaque, partially transparent, patchy) and geometries can be invoked to explain the spectral curvature. 

In the soft-Comptonisation model, while Comptonisation from the hot, primary corona produces the hard X-ray power law at energies above $\sim2\kev$, the soft excess can also be produced from Compton up-scattering of UV seed photons by a secondary, cooler corona. This secondary corona exists as a thin layer above the accretion disc, and is optically thick (e.g. \citealt{Done+2012}). The resulting spectrum features a smooth soft excess due to the blending of the two power law components. This feature is seen in some AGN including Ark 120 (e.g. \citealt{Vaughan+2004,Porquet+2018}), Mrk 530 (\citealt{Ehler+2018}), and Zw 229.015 (\citealt{zw229}). A separate source of emission is then required to explain the presence of \feka\ lines and other emission features. 

In the blurred reflection model, some of the photons emitted from the corona will shine on the inner accretion disc, producing a reflection spectrum (e.g. \citealt{Fabian+2005}). When the coronal photons strike the accretion disc, they are absorbed by atoms in the top layer, which produces absorption features. As the atoms de-excite, they undergo fluorescence, which produces a multitude of emission lines. This includes an emission line at $6.4\kev$ associated with \feka\ emission. As the material in the innermost regions of the accretion disc rotates, it is subjected to the effects of general relativity, which broaden the intrinsically narrow emission lines and produces an excess of photons below $2\kev$. The resulting X-ray spectrum can show different levels of contributions from the reflection and primary emission components.  If the X-ray emitting corona is sufficiently close to the black hole, more photons will be pulled back towards the black hole itself due to its extreme gravity in a process known as light bending (e.g. \citealt{Miniutti+2004}), producing a reflection dominated spectrum.  On the other hand, if the corona is moving away from the central region, the primary emission may be beamed towards the observer and produce a power law dominated spectrum. In a more typical system, the isotropic emission from the corona implies comparable contributions from the reflection and power law components, and reflection fraction (R) values are $\simeq1$.

Mrk 478 (PG1440+356; z=0.079) is a luminous, nearby NLS1 (\citealt{Zoghbi+2008}). The source appears in some larger AGN surveys given its Palomar-Green and Markarian classifications (e.g. \citealt{Boroson+1992,Leighly1999,Porquet+2004,Grupe+2010}). \cite{Leighly1999} analyzed the \asca\ (\citealt{ascains}) data as part of a large sample of NLS1s, and found that the spectrum can be characterised with a power law of slope $\sim2$, a black body with temperature of $\sim120\eV$, and a weak, broad ($0.9\kev$) feature at $\sim6\kev$ attributed to \feka\ emission. Mrk 478 was not detected in the \swift\ BAT 105 month catalogue (\citealt{Oh+2018}), indicating a low X-ray flux at high energies.

An 80ks \chandra\ (\citealt{chandrains}) LETG exposure was obtained in 2000. The data were well fitted with a power law and Galactic absorption, and no significant narrow emission or absorption lines were found, suggesting a lack of obscuring material in the vicinity (\citealt{Marshall+2003}). Comparing this result with a near simultaneous \sax\ Medium Energy Concentrator Spectrometer (MECS; \citealt{saxins}) observation, \cite{Marshall+2003} concluded that the soft excess is likely produced by Comptonisation of thermal emission from the accretion disc by a thin corona on top of the disc. 

Mrk 478 was also the subject of four short ($\sim20\ks$) \xmm\ (\citealt{xmmins}) observations over 13 months between 2001 and 2003. \cite{Guainazzi2004} found changes in $0.35-15\kev$ flux by a factor of $\sim6$ and that most spectral changes occurred above $3\kev$. They also suggested that the emission from the AGN could be explained with a double Comptonisation scenario. The soft and hard-band light curves from these observations were examined, and it was found that on short (hourly) timescales, the hard and soft bands varied together, while on longer timescales the soft flux varies more dramatically. \cite{Guainazzi2004} also noted that the while the soft X-ray flux is variable, the spectral shape does not significantly change between observations.

\cite{Zoghbi+2008} studied the same \xmm\ data, noting the same large flux variations as \cite{Guainazzi2004}. They also noted that the ratio of hard to soft photons was not dependent on changes in flux. In contrast to \cite{Marshall+2003}, however, \cite{Zoghbi+2008} concluded that the spectrum is dominated by a blurred reflection component, which accounted for  $\sim90$ per cent of the total observed X-ray emission in each observation. They found that a highly blurred spectrum originating from the inner region of the accretion disc around a Kerr black hole produced an excellent fit to the smooth soft excess. Curiously, they found that the model also requires a sub-solar iron abundance.

In this work, a multi-epoch analysis is conducted using all available data from \xmm\ and \suzaku\ from 2001 to 2017. Only the average properties of these spectra are considered. For the longest exposure, obtained in 2017, flux-resolved spectral and timing analysis will be presented in a future work (Waddell \et\ in prep). A variety of spectral models are compared to attempt to explain the X-ray emission and describe the long-term spectral variations. In Section~\ref{sect:data}, the observations used in the work are tabulated, and data reduction and modelling techniques are outlined. In Section~\ref{sect:var}, long and short term light curves are discussed. Section~\ref{sect:model} outlines the different models used to fit the spectra (partial covering, soft-Comptonisation, and blurred reflection). Section~\ref{sect:discussion} presents a discussion of the results, and final conclusions are drawn in Section~\ref{sect:conclusion}.

%%%%%%%%%%%%%%%%%%%%%%%%%%%%%%%%%%%%%%%%%%%%%%%%%%%%%%%%%%%%%%%%%%%%%%%%%%%%%%%%%%%%%%%%%%%%%%%%%

\section{Observations and Data Reduction}
\label{sect:data}

\subsection{\xmm}
\label{sect:xmm}
Mrk 478 has been observed a total of five times between 2001 and 2017 with \xmm\ (\citealt{xmmins}) and all observations are shown in Table~\ref{tab:obs}. Spectra are referred to as XMM1 - XMM5. The \xmm\ Observation Data Files (ODFs) were processed to produce calibrated event lists using the \xmm\ Science Analysis System ({\sc sas v15.0.0}). Light curves created from these event lists were then checked for evidence of background flaring. This was significant in XMM1 and XMM3, so good time intervals (GTI) were created and applied. Further corrections were applied to XMM3 after examining the $0.3-10\kev$ background light curve and finding high count rates in the second half of the observation. Only the first $12\ks$ are used for this observation, and this spectrum is only considered up to $9.0\kev$. 

Evidence for pile-up on the order of $10$ per cent was also found in XMM1. Attempting to correct for this effect resulted in a loss of $\sim60$ per cent of detected photons, but did not noticeably affect the shape of the spectrum, so no corrections were made.

Source photons were extracted from a circular region with a $35\arcs$ radius centred on the source, and background photons were extracted from an off-source circular region with a $50\arcs$ radius on the same CCD. Single and double events were selected for the EPIC-pn (\citealt{pnins}) detector products, while single through quadruple events were selected for the EPIC-MOS (\citealt{mosins}) detectors. The {\sc sas} tasks {\sc rmfgen} and {\sc arfgen} were used to generate response files. Light curves using the same source and background regions were extracted, and background corrected using {\sc epiclccorr}.

\cite{Zoghbi+2008} determined that the spectra from MOS1 and MOS2 were comparable to the pn data for observations XMM1 through XMM4. We also compared the spectra from MOS1 and MOS2 to the pn spectra from XMM5, and determined these to be consistent. However, the effective areas of MOS1 and MOS2 are smaller than those of the pn detector, resulting in fewer counts from these instruments. Therefore, for simplicity, only data from the pn detector are considered for the remainder of this work.

%%%%%%%%%%%%%%%%%%%%%%%%%%%%%%%%%%%%%%%%%%%%%%%%%%%%%%%%%%%%%%%%%%%%%%%%%%%%%%%%%%%%%%%%%%%%%%%%%

\begin{table*}
	\centering
	\begin{tabular}{c c c c c c c c}
	    \hline
	    (1) & (2) & (3) & (4) & (5) & (6) & (7) & (8)\\
	    Observatory & Observation ID & Name & Start Date & Duration & Exposure & Counts & Energy Range\\
	    & & & (yyyy-mm-dd) & [s] & [s] & & [keV]\\
	    \hline
	    \xmm\ EPIC-pn & 0107660201 & XMM1 & 2001-12-23 & 32616 & 20390 & 113966 & 0.3-10.0\\
	    & 0005010101 & XMM2 & 2003-01-01 & 27751 & 17230 & 94079 & 0.3-10.0\\
	    & 0005010201 & XMM3 & 2003-01-04 & 29435 & 8371 & 39293 & 0.3-9.0\\
	    & 0005010301 & XMM4 & 2003-01-07 & 26453 & 18180 & 49822 & 0.3-10.0\\
	    & 0801510101 & XMM5 & 2017-06-30 & 135300 & 93710 & 291574 & 0.3-10.0\\
	    \hline
	    \suzaku\ XIS0+3 & 706041010 & SUZ & 2011-07-14 & 170600 & 85323 & 42336 & 0.7-8.0\\
	    \hline
	\end{tabular}
	\caption{Observations used for spectral modelling. Observations are named sequentially in column (3), and these names will be used for the remainder of this work. Column (6) shows the exposure after GTI's were applied, and column (8) gives the energy range used in spectral modelling. For \suzaku\, the combined counts from XIS0 and XIS3 are given (column 7).}
	\label{tab:obs}
\end{table*}

%%%%%%%%%%%%%%%%%%%%%%%%%%%%%%%%%%%%%%%%%%%%%%%%%%%%%%%%%%%%%%%%%%%%%%%%%%%%%%%%%%%%%%%%%%%%%%%%%

Data from the RGS instrument (\citealt{rgsins}) were also obtained during this time. The data were reduced using the {\sc sas} task {\sc rgsproc}. First order spectra were obtained from RGS1 and RGS2, and were merged using the task {\sc rgscombine}. 

Optical Monitor (OM) (\citealt{omins}) imaging mode data were obtained during all observations, including data from UVW2 in all epochs, and UVW1 in all epochs except XMM5. Data were reduced using the {\sc sas} task {\sc omichain}, from which the average count rate in each filter was obtained. Count rates were then converted to fluxes using standard conversion tables. These were then corrected for reddening using $E(B-V) = 0.011$ (\citealt{Willingale+2013}).

%%%%%%%%%%%%%%%%%%%%%%%%%%%%%%%%%%%%%%%%%%%%%%%%%%%%%%%%%%%%%%%%%%%%%%%%%%%%%%%%%%%%%%%%%%%%%%%%%
%%%%%%%%%%%%%%%%%%%%%%%%%%%%%%%%%%%%%%%%%%%%%%%%%%%%%%%%%%%%%%%%%%%%%%%%%%%%%%%%%%%%%%%%%%%%%%%%%

\subsection{\suzaku}
\label{sect:suz}
\suzaku\ (\citealt{suzins}) observed Mrk 478 in HXD (Hard X-ray Detector) nominal mode using the front illuminated (FI) CCDs XIS0 and XIS3, the back illuminated (BI) CCD XIS1, and the HXD-PIN detector. Cleaned event files from the HXD-PIN detector were processed using the tool {\sc hxdpinxbpi}, yielding $72\ks$ of good time exposure. After considering both the instrumental and cosmic ray backgrounds, the observation resulted in a null detection, with a detection significance of only a few per cent. 

Cleaned event files from the FI and BI CCDs were used for the extraction of data products in {\sc xselect v2.4d}. For each instrument, source photons were extracted using a $240\arcs$ region centred around the source, while background photons were extracted from a $180\arcs$ off-source region. Calibration zones located in the corners of the CCDs were avoided in the background extraction. Response matrices for each detector were generated using the tasks {\sc xisrmfgen} and {\sc xissimarfgen}. The XIS0 and XIS3 detectors were first checked for consistency. Spectra were then merged using the task {\sc addascaspec}. The resulting FI spectrum was found to be background dominated at $E > 8\kev$, so only data from $0.7-8.0\kev$ are considered. The regions $1.72-1.88\kev$ and $2.19-2.37\kev$ are also excluded from spectral analysis because of calibration uncertainties (\citealt{Nowak+2011}). The merged FI data were checked for consistency against the BI spectrum and found to be consistent. Only the FI data are presented for simplicity.

%%%%%%%%%%%%%%%%%%%%%%%%%%%%%%%%%%%%%%%%%%%%%%%%%%%%%%%%%%%%%%%%%%%%%%%%%%%%%%%%%%%%%%%%%%%%%%%%%%%%%%%%%%%%%%%%%%%%%%%%%%%%%%%%%%%%%%%%%%%%%%%%%%%%%%%%%%%%%%%%%%%%%%%%%%%%%%%%%%%%%%%%%%%%%%%%%%

\subsection{\swift}
\label{sect:swift}
Mrk 478 has been observed with \swift\ (\citealt{swiftins}) XRT (\citealt{xrtins}) a total of fifteen times between 2006 and 2017, with exposures ranging from $0.1-5\ks$. Data products are extracted using the web tool \swift\ XRT data products generator{\footnote {http://www.swift.ac.uk/user\_objects/}} (\citealt{Evans+2009}) and the average count-rate from each observation is presented. The averaged, background subtracted spectrum created using all observations is modelled with a power law + black body, which over-fits the data ($\redchi = 0.74$). These parameters are used to obtain the $2-10\kev$ flux based on the count rate at each epoch using the {\sc webpimms}{\footnote{https://heasarc.gsfc.nasa.gov/cgi-bin/Tools/w3pimms/w3pimms.pl}} tool. The spectrum has poor signal-to-noise due to the short exposure, and is not used in further spectral modelling.

%%%%%%%%%%%%%%%%%%%%%%%%%%%%%%%%%%%%%%%%%%%%%%%%%%%%%%%%%%%%%%%%%%%%%%%%%%%%%%%%%%%%%%%%%%%%%%%%%
%%%%%%%%%%%%%%%%%%%%%%%%%%%%%%%%%%%%%%%%%%%%%%%%%%%%%%%%%%%%%%%%%%%%%%%%%%%%%%%%%%%%%%%%%%%%%%%%%

\subsection{Spectral Fits}
\label{sect:fits}
All spectral fits are performed using {\sc xspec} v12.9.0n (\citealt{xspec}) from {\sc heasoft} v6.26. Spectra from the \xmm\ EPIC-pn and \suzaku\ FI detectors have been grouped using optimal binning (\citealt{optbin}), using the {\sc ftool} {\sc ftgrouppha}. All spectra are background subtracted. Model fitting was done using C-statistic (\citealt{Cash1979}). All model parameters are reported in the rest frame of the source, but figures are shown in the observed frame. The Galactic column density in the plane of Mrk 478 is kept frozen at $1.08\times10^{20}\cm^{-2}$ (\citealt{Willingale+2013}) for all models, and elemental abundances used for this parameter only are taken from \cite{Wilms+2000}.

All models were originally fit with parameters free to vary between all data sets. However, because of the low exposure, high background, and overall poor data quality of XMM3, many parameters could not be well constrained at this epoch. Consequently, all parameters for XMM3 are linked to those of XMM1, since the two spectra are almost identical in shape (see Fig.~\ref{fig:eeuf_plext}). A constant is left free to vary between the data sets to account for the small flux differences. This technique does not result in any significant change in fit quality or parameters for any model.

To determine parameter errors from spectral fits, Monte Carlo Markov Chain (MCMC) techniques are employed using {\sc xspec\_emcee}{\footnote {Made available by Jeremy Sanders (http://github.com/jeremysanders/xspec\_emcee)}}. After obtaining the best-fitting parameters, MCMC calculations are run. The Goodman-Weare (\citealt{GW+2010}) algorithm is used, and each chain is run with at least twice as many walkers (64 walkers) as there are free parameters to ensure sufficient sampling. Chain lengths are set to be at least 10000. A burn-in phase of 1000 is selected to ensure that no bias is introduced based on the starting parameters. All errors on parameters are quoted at the $90$ per cent confidence level.

%%%%%%%%%%%%%%%%%%%%%%%%%%%%%%%%%%%%%%%%%%%%%%%%%%%%%%%%%%%%%%%%%%%%%%%%%%%%%%%%%%%%%%%%%%%%%%%%%
%%%%%%%%%%%%%%%%%%%%%%%%%%%%%%%%%%%%%%%%%%%%%%%%%%%%%%%%%%%%%%%%%%%%%%%%%%%%%%%%%%%%%%%%%%%%%%%%%
%%%%%%%%%%%%%%%%%%%%%%%%%%%%%%%%%%%%%%%%%%%%%%%%%%%%%%%%%%%%%%%%%%%%%%%%%%%%%%%%%%%%%%%%%%%%%%%%%

\section{Characterising the Variability}
\label{sect:var}

Light curves for the $2001-2003$ (XMM1--XMM4) \xmm\ observations in the $0.2-10\kev$ band using $200\s$ bins were presented in \cite{Zoghbi+2008}. Here, light curves for all the \xmm\ observations (XMM1--XMM5) are created in the $0.3-10\kev$ band, to match energy ranges used in spectral modelling, using $1000\s$ bins. Additionally, the \suzaku\ light curve is created between $0.7-10\kev$, using a bin size of $5760\s$ corresponding to the \suzaku\ orbit. 

The shapes of light curves are similar to those of \cite{Zoghbi+2008}. Variations in the count rate are on the order of $15-20$ per cent about the average during the short exposures (< $20\ks$, XMM1 through XMM4), and on the order of $40$ per cent for the longer XMM5. The \suzaku\ light curve shows significant deviations of $\sim80$ per cent from the mean, over the course of days.

Normalized hardness ratios were presented by \cite{Zoghbi+2008} for XMM1 and XMM4 using $S=0.3-2\kev$ and $H=3-10\kev$, with $200\s$ binning. The ratios were found to be consistent with a constant over the course of the observations. Hardness ratios are calculated here using $HR = H/S$, where $S=0.3-2\kev$ and $0.7-2\kev$ for \xmm\ and \suzaku, respectively, and $H=2-10\kev$. The same binning as in the light curves is used for the hardness ratios. For all observations, variations in $HR$ are on the order of $10-20$ per cent. Given the modest hardness ratio variability within each observation, the average SUZ and XMM5 spectra are used in the analysis, despite their longer length.

\begin{figure}
	\includegraphics[width=\columnwidth]{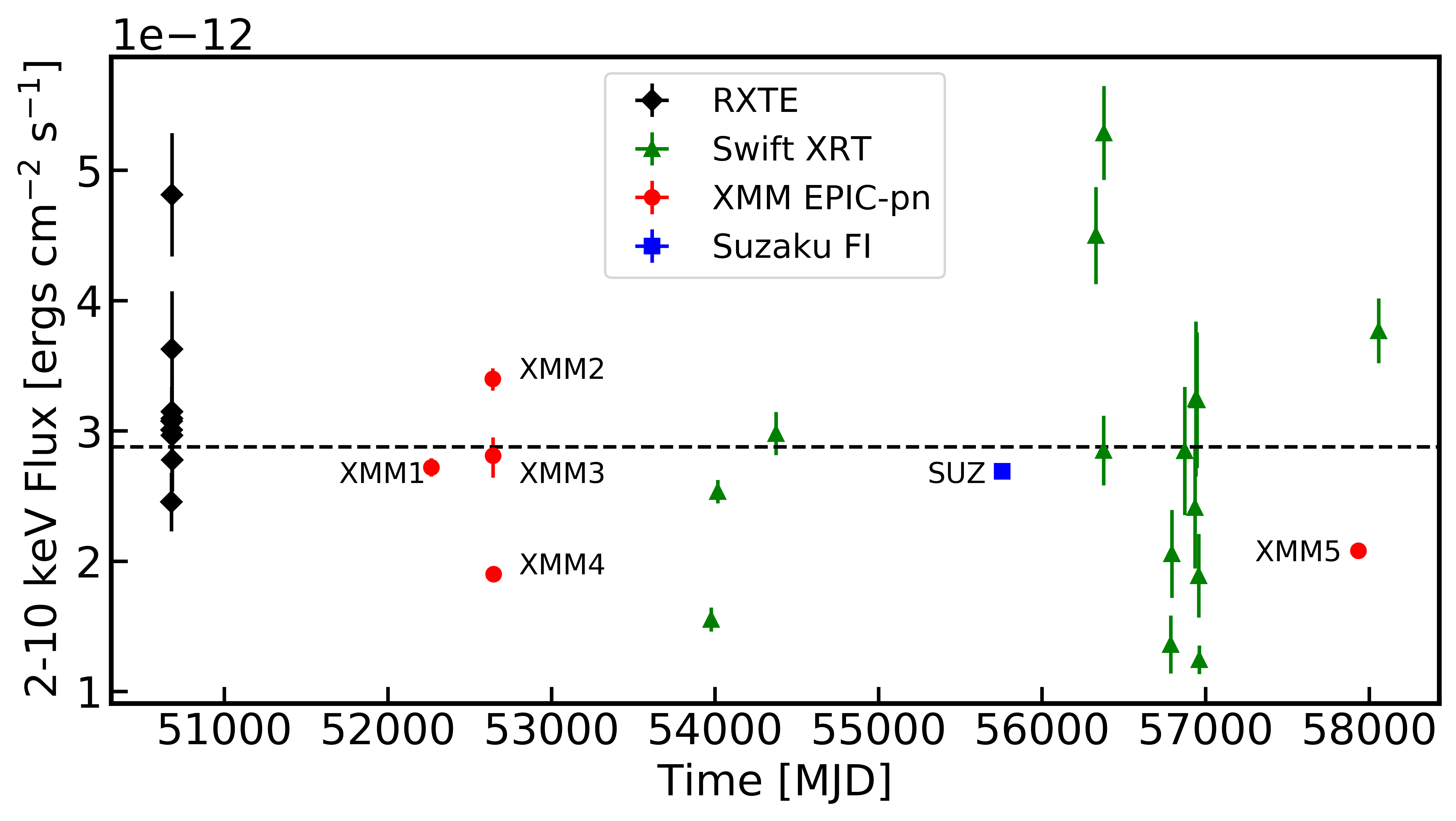}
	\caption{Long term light curve between $1997-2017$ for Mrk 478 with data from \rxte\ (black diamonds), \swift\ (green triangles), \xmm\ (red circles) and \suzaku\ (blue squares). The $2-10\kev$ flux is shown on the y-axis, and the average flux is shown as a dashed black line.}
	\label{fig:lcdiv}
\end{figure}

The $2-10\kev$ light curve for Mrk 478 spanning from 1997 to 2017 is presented in Fig.~\ref{fig:lcdiv} making use of data from \swift, \xmm, \suzaku, and the Rossi X-ray Timing Explorer (\rxte). The \rxte\ data were taken from the \rxte\ AGN Timing \& Spectral Database{\footnote {https://cass.ucsd.edu/$\sim$rxteagn/}} (\citealt{Breedt+2009,Rivers+2013}). A total of nine short exposures were taken between August 16-21, 1997, lasting between $1-10\ks$ each. 

The average flux between $1997-2017$ is approximately $2.88\times10^{-12}\ergpscmps$. Between the dimmest and brightest observations, variations by a factor of $\sim5$ are seen, and deviations from the mean are on the order of $80$ per cent at the extremes. By comparing each observation to the average flux, it is apparent that XMM2 is found at a slightly higher than average flux state, XMM1, XMM3 and SUZ are at average flux states, and XMM4 and XMM5 are at a dimmer flux level. Overall, the variability is less significant than is seen in more extreme NLS1s, like Mrk 335 (e.g. \citealt{Gallo+2019m335,Wilkins+2015}) and IRAS 13224-3809 (e.g. \citealt{Alston+2019}). 

To assess the UV-to-X-ray shape at each epoch, the hypothetical power law between $2500$ angstrom and $2\kev$ ($\alpha_{\rm ox}$; \citealt{Tananbaum+1979}) is calculated. Comparing the measured $\alpha_{\rm ox}$ to the expected value given the UV luminosity ($\alpha_{\rm ox}(L_{2500}$); \citealt{Vagnetti+2013}) reveals the X-ray weakness parameter ($\Delta \alpha_{\rm ox} = \alpha_{\rm ox} - \alpha_{\rm ox}(L_{2500}$)), where negative values indicate X-ray weak sources and positive values imply X-ray strong sources relative to the UV luminosity. 

The resulting $\Delta \alpha_{\rm ox}$ values range from $-0.07$ for XMM2 to $-0.17$ for XMM4, but are comparable within uncertainties. All values are negative, perhaps implying that all epochs are slightly X-ray weak. \cite{Gallo2006} suggests that more X-ray weak sources (negative $\Delta \alpha_{\rm ox}$ values) are likely to be accompanied with increased spectral complexity and are more reflection dominated or more highly absorbed. However, the distribution on the expected $\alpha_{\rm ox}$ is large (\citealt{Vagnetti+2013}), so all values measured for Mrk 478 agree with one another and with $\Delta \alpha_{\rm ox}\approx0$. This implies that Mrk 478 is in an X-ray normal state (e.g. \citealt{Gallo2006}). Therefore, extreme spectra that are  highly absorbed or highly dominated by reflection are not expected.

%%%%%%%%%%%%%%%%%%%%%%%%%%%%%%%%%%%%%%%%%%%%%%%%%%%%%%%%%%%%%%%%%%%%%%%%%%%%%%%%%%%%%%%%%%%%%%%%%

%%%%%%%%%%%%%%%%%%%%%%%%%%%%%%%%%%%%%%%%%%%%%%%%%%%%%%%%%%%%%%%%%%%%%%%%%%%%%%%%%%%%%%%%%%%%%%%%%

Interestingly, both the short and long term light curves, as well as the calculated $\alpha_{\rm ox}$ values, can be interpreted in similar fashion. First, all are suggestive of a relatively X-ray "normal" spectrum, without extreme flux or spectral changes. This is true on timescales of days (within observations) and throughout the 20 years of observation. This suggests a lack of excessive spectral complexity produced by complicated absorption or blurred reflection. Secondly, the variability on short and long time scales is likely simple, produced mainly by normalisation changes. The multi-epoch analysis should provide good constraint on parameters that are not expected to vary between observations.

%%%%%%%%%%%%%%%%%%%%%%%%%%%%%%%%%%%%%%%%%%%%%%%%%%%%%%%%%%%%%%%%%%%%%%%%%%%%%%%%%%%%%%%%%%%%%%%%%
%%%%%%%%%%%%%%%%%%%%%%%%%%%%%%%%%%%%%%%%%%%%%%%%%%%%%%%%%%%%%%%%%%%%%%%%%%%%%%%%%%%%%%%%%%%%%%%%%
%%%%%%%%%%%%%%%%%%%%%%%%%%%%%%%%%%%%%%%%%%%%%%%%%%%%%%%%%%%%%%%%%%%%%%%%%%%%%%%%%%%%%%%%%%%%%%%%%

\section{Spectral Modelling}
\label{sect:model}

\subsection{Spectral Characterisation}
\label{sect:phen}

\begin{figure}
	\includegraphics[width=\columnwidth]{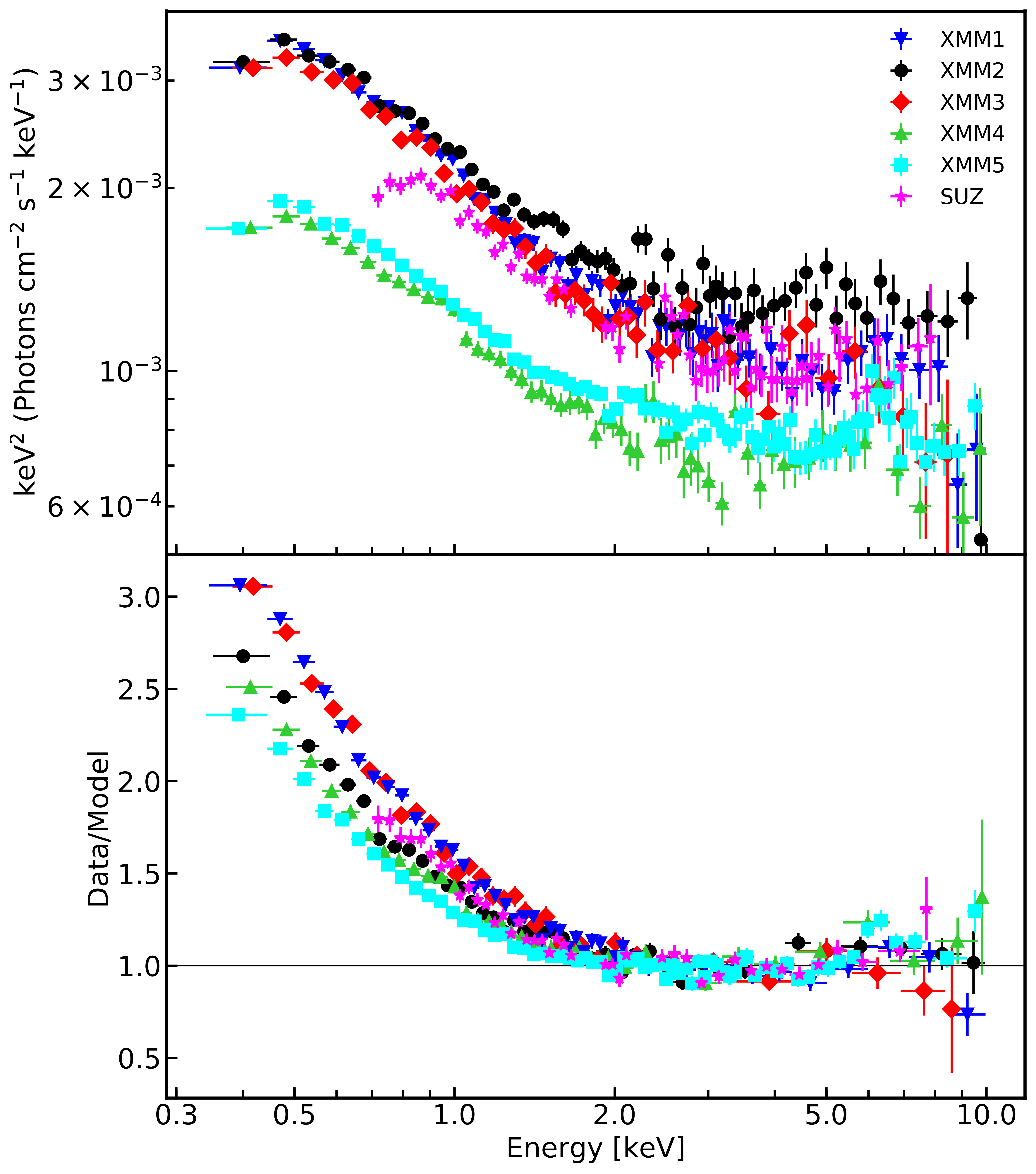}
	\caption{\textbf{Top panel:} Data from each epoch unfolded against a power law with $\Gamma=0$. The shape of all spectra are comparable, with XMM4 and XMM5 being dimmer than other epochs. \textbf{Bottom panel:} Residuals from a power law fit from $2-4$ and $7-10\kev$, extrapolated over the $0.3-10\kev$ band. All data sets are fit with the average power law, and scaled by a constant to account for flux variations between observations. A strong soft excess below $2\kev$ and excess residuals between $5-7\kev$ are evident in all data sets.}
	\label{fig:eeuf_plext}
\end{figure}

To obtain an initial assessment of differences between spectra, data from each epoch are unfolded against a power law with $\Gamma=0$. This allows data from different instruments to be compared on the same plot. The result is shown in the top panel of Fig.~\ref{fig:eeuf_plext}, and data have been re-binned in {\sc xspec} for clarity. The colours and symbols used in Fig.~\ref{fig:eeuf_plext} are adopted throughout the remainder of this work. The shape of the spectra do vary between observations, however, most of the changes are likely attributable to normalisation changes between lower flux level spectra (XMM4 and XMM5) and higher levels (XMM1, XMM2, XMM3 and SUZ), in agreement with the findings of \cite{Guainazzi2004}. 

%%%%%%%%%%%%%%%%%%%%%%%%%%%%%%%%%%%%%%%%%%%%%%%%%%%%%%%%%%%%%%%%%%%%%%%%%%%%%%%%%%%%%%%%%%%%%%%%%

%%%%%%%%%%%%%%%%%%%%%%%%%%%%%%%%%%%%%%%%%%%%%%%%%%%%%%%%%%%%%%%%%%%%%%%%%%%%%%%%%%%%%%%%%%%%%%%%%

In an attempt to characterise the \xmm\ and \suzaku\ spectra, the data are fit from $2-10\kev$, excluding the $4-7\kev$ band where iron emission is typically detected, with a single average power law. A constant factor is applied to account for changes in flux between the different spectra. The model is then extrapolated over the full usable energy range for each data set and the ratio (data/model) is shown in the bottom panel of Fig.~\ref{fig:eeuf_plext}. A smoothly rising soft excess below $\sim2\kev$ is evident in all data sets, as well as some evidence for excess residuals in the $4-7\kev$ band. This also shows some spectral variability between observations, as the spectra appear softer when brighter, which is typical of X-ray binaries and AGN.

To model the soft excess, a black body is added to the spectrum from each epoch. The soft excess can be characterised by a black body with temperatures of $\sim90\eV$ at all epochs, and a steep photon index of $\sim2.4$. However, the data are not well fitted by the model, with significant curvature around $1\kev$ and in the $4-10\kev$ band. Replacing the black body component with a secondary power law improves the fit, but is still unable to explain the overall shape and does not fit the spectrum well. For either model, adding a Gaussian emission feature results in a good fit to the residuals in the $4-7\kev$ band. The feature has a best fit energy of $\sim6.7\kev$ and a width of $\sim0.5\kev$. This may indicate a broad \feka\ line profile, however the high best fit energy may also indicate the presence of ionised iron emission lines at $6.7\kev$ (\fexxv) and $6.97\kev$ (\fexxvi). More physical models will be examined to explain the observed spectrum.

The RGS data from the long \xmm\ exposure in 2017 (XMM5) are also modelled with a power law and Galactic absorption. The best fit photon index for the power law component is $\Gamma=2.83\pm0.03$. No strong emission or absorption lines are detected in the $0.3-2\kev$ range. The lack of any strong emission or absorption features in the RGS data are consistent with the \chandra\ LETG observation in 2000 (\citealt{Marshall+2003}).

%%%%%%%%%%%%%%%%%%%%%%%%%%%%%%%%%%%%%%%%%%%%%%%%%%%%%%%%%%%%%%%%%%%%%%%%%%%%%%%%%%%%%%%%%%%%%%%%%

\begin{figure*}
	\includegraphics[width=\textwidth]{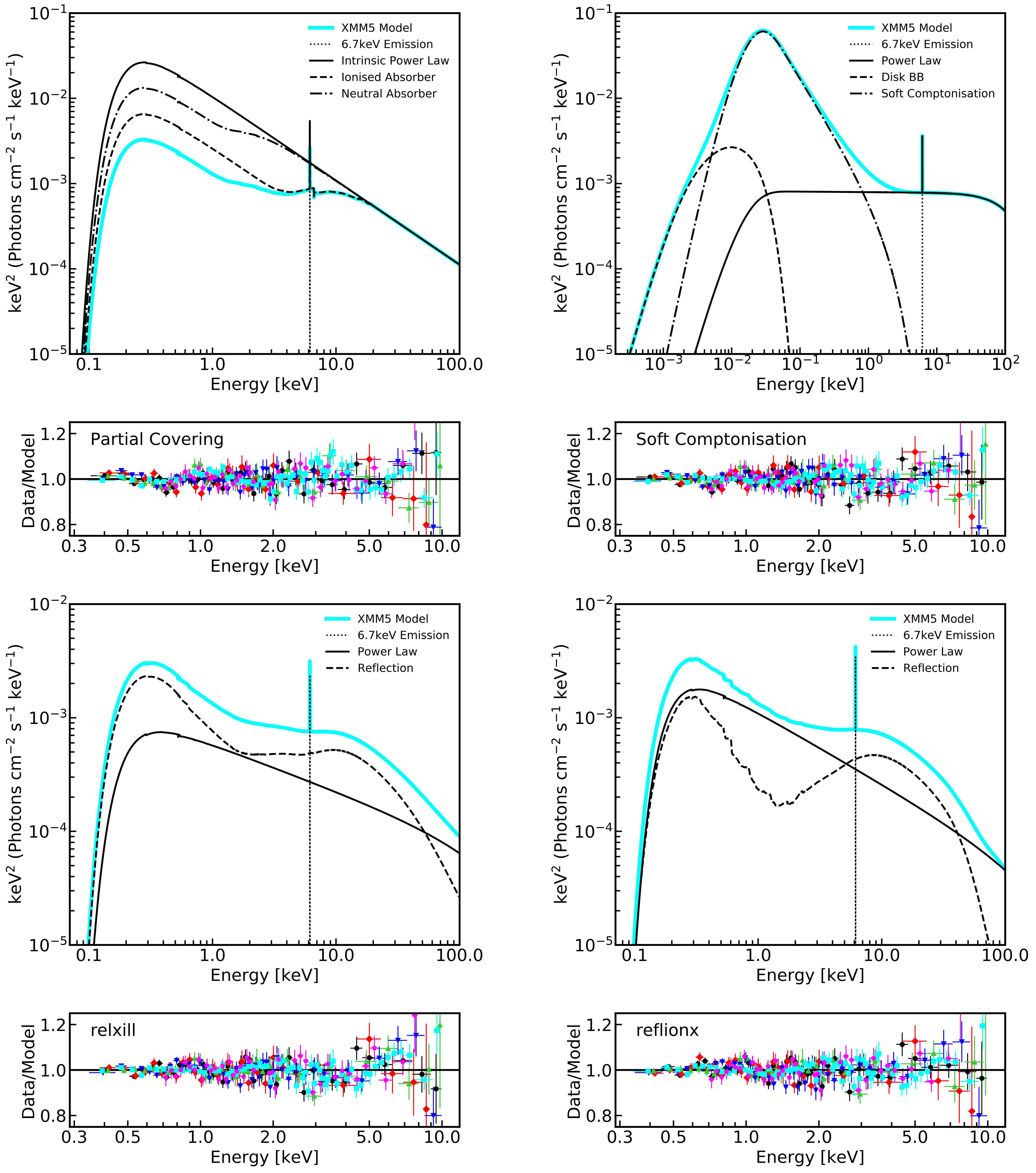}
	\caption{Best-fit model and residuals for physical models in this work. Models are only shown for XMM5 for clarity. \textbf{Top left:} Model components and residuals for the partial covering scenario. Contributions from the intrinsic power law, ionised absorber, neutral absorber and total model are shown. Residuals (data/model) are shown in the panel below, displaying significant curvature at high and low energies. \textbf{Top right:} Model components and residuals for the double Comptonisation model. Effects of Galactic absorption have been removed for clarity, and the axes have been modified accordingly. Contributions from the disc black body, soft-Comptonisation, and hard Comptonisation are shown. \textbf{Bottom left:} Model components and residuals for the blurred reflection model using {\sc relxill}. The reflection and power law components are shown, and the intrinsic spectrum is clearly reflection dominated. \textbf{Bottom right:} Model components and residuals for the blurred reflection model using {\sc reflionx}. The reflection and power law components are again shown, and the resulting spectrum is not dominated by reflection.}
	\label{fig:model}
\end{figure*}

%%%%%%%%%%%%%%%%%%%%%%%%%%%%%%%%%%%%%%%%%%%%%%%%%%%%%%%%%%%%%%%%%%%%%%%%%%%%%%%%%%%%%%%%%%%%%%%%%
%%%%%%%%%%%%%%%%%%%%%%%%%%%%%%%%%%%%%%%%%%%%%%%%%%%%%%%%%%%%%%%%%%%%%%%%%%%%%%%%%%%%%%%%%%%%%%%%%

\subsection{Partial Covering}
\label{sect:pc}

Partial covering has been used to successfully describe the soft excess and high energy curvature in NLS1 X-ray spectra (e.g. \citealt{Tanaka+2004, Gallo+2015}). To find a partial covering model to explain the observed spectral shape and variability, all data sets are first modelled with a single, constant power law modified by neutral absorption ({\sc zpcfabs}). The redshift of the absorber is the same as the host galaxy. To describe the spectral changes, the column density and covering fraction of the absorber are free to vary between data sets. This gives $C = 1998$ for 557 degrees of freedom (dof). The model underestimates the data at $\sim4\kev$ and between $7-10\kev$. Allowing the photon index to vary between epochs improves the fit; $\Delta C = 129$ for 4 additional free parameters, but $\Gamma$ remains comparable ($\sim2.9$) for all data sets. 

To explain the residuals left by the single partial covering model, a second {\sc zpcfabs} component is added. The column density and covering fraction are again left free to vary between observations. This improves the fit, with $\Delta C = 1244$ for 10 additional free parameters. The analysis suggests two distinct absorbers, both with very different covering fractions and column densities. One absorber has a column density of $\sim4-7\times10^{22}$ cm$^{-2}$ and covering fraction of $\sim0.4$, while the other has column densities of $\sim80\times10^{22}$ cm$^{-2}$ and high covering fraction of $\sim0.7$. This higher density absorber produces a deep \feka\ edge, which fits the data much better at high energies; however, some excess residuals are still visible at low energies in all data sets. 

One neutral component is then replaced with an ionised absorber ({\sc zxipcf}). This model includes the column density and covering fraction of the absorber, as well as an ionisation parameter ($\xi = 4\pi F/n$, where F is the illuminating flux and n is the hydrogen number density of the absorber). The redshift of this absorber is again fixed to that of the host galaxy. Fitting the data gives $C = 676$ for 544 dof. Although the ionisation parameter is low, the use of {\sc zxipcf} results in much lower column densities for the secondary absorber, causing the large change in fit statistic. Replacing the remaining neutral absorber with a second ionised absorber does not improve the fit, so the combination of one neutral and one ionised absorber is maintained. 

To describe the variability, various parameter combinations are examined. The best fit is obtained when the ionisation parameter is linked between epochs, but the column density and covering fractions of both absorbers are left free to vary. The slope and normalisation of the power law component are also kept linked between epochs. 

A close examination of the residuals reveals some emission in the $6-7\kev$ range. A narrow emission line is added. The width is kept fixed at $1\eV$, and the line energy and normalisation are left free to vary, but kept linked between data sets. This again improves the fit, $\Delta C = 9$ for 2 additional free parameters. No signatures of a $6.4\kev$ emission feature are detected. The best fit energy is at $\sim6.7\kev$, indicating the presence of ionised iron emission as suggested in Section~\ref{sect:phen}.

%%%%%%%%%%%%%%%%%%%%%%%%%%%%%%%%%%%%%%%%%%%%%%%%%%%%%%%%%%%%%%%%%%%%%%%%%%%%%%%%%%%%%%%%%%%%%%%%%

\begin{figure}
	\includegraphics[width=\columnwidth]{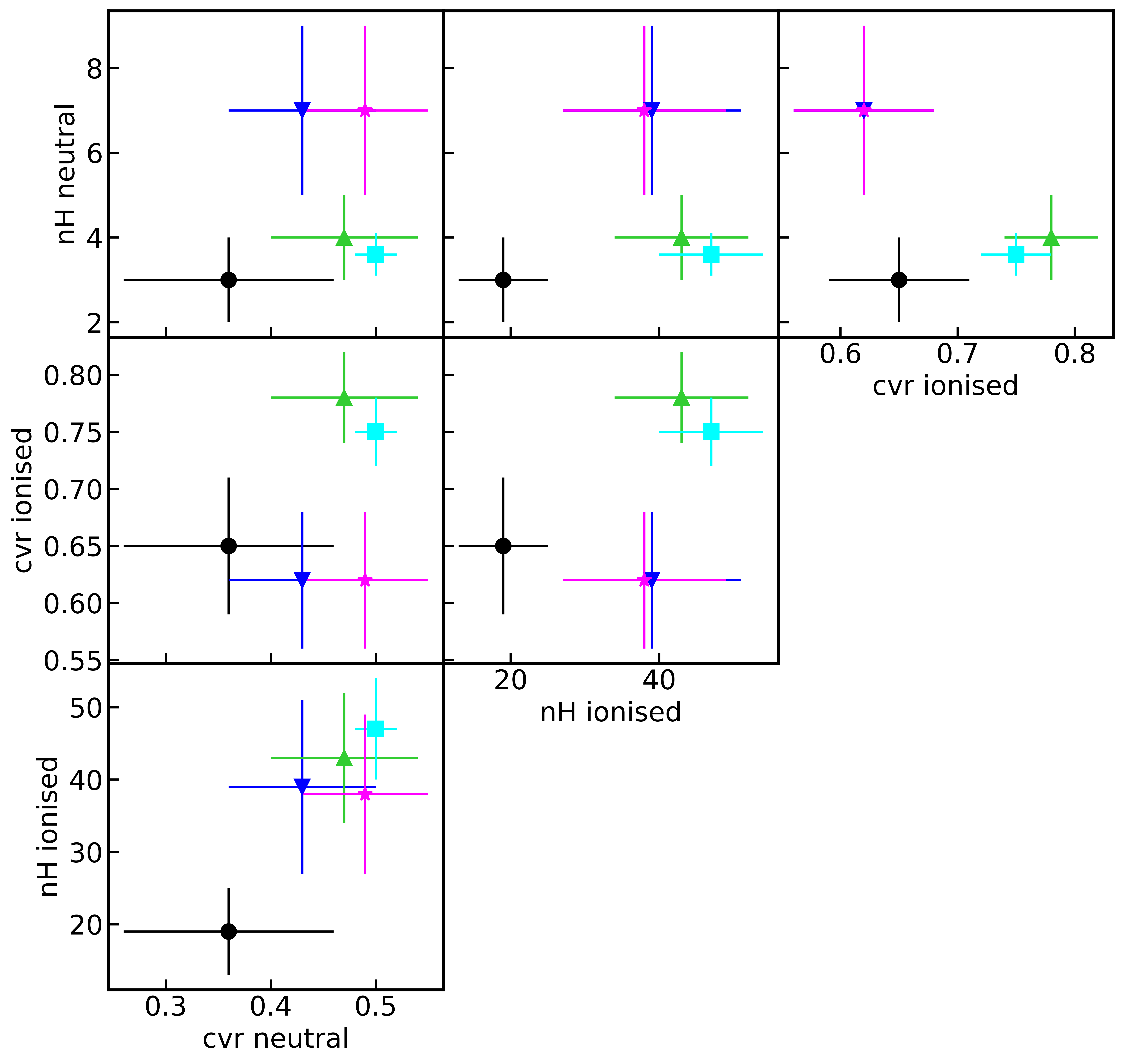}
	\caption{Correlations between best fit parameters using the partial covering model. While no linear trends are observable between data sets, the column density and covering fraction are higher for the dimmer flux XMM4 and XMM5 than for other data sets.}
	\label{fig:pcfcorr}
\end{figure}

%%%%%%%%%%%%%%%%%%%%%%%%%%%%%%%%%%%%%%%%%%%%%%%%%%%%%%%%%%%%%%%%%%%%%%%%%%%%%%%%%%%%%%%%%%%%%%%%%

The best fit parameters and MCMC errors for this final model are presented in Table~\ref{tab:covering}, and the model and residuals are shown in the top-left corner of Fig.~\ref{fig:model}. Parameter correlations for the best fit model are presented in Fig.~\ref{fig:pcfcorr}. The intrinsic power law (shown in Fig.~\ref{fig:model}) is extremely steep, with $\Gamma = 2.99\pm0.02$. The ionisation parameter of {\sc zxipcf} is low, poorly constrained, and consistent with neutral material. The neutral absorber ({\sc zpcfabs}) has a low column density and low covering fraction, and displays only limited variability between epochs. 

The ionised absorber displays more significant variability; in particular, the covering fraction and density are both higher for the dimmer XMM4 and XMM5 observations, and are lower for the brightest data set (XMM2). The higher overall densities and covering fractions suggest that this component is driving the shape and variability of the observed spectra rather than changes to the intrinsic power law emission or neutral absorption. 

%%%%%%%%%%%%%%%%%%%%%%%%%%%%%%%%%%%%%%%%%%%%%%%%%%%%%%%%%%%%%%%%%%%%%%%%%%%%%%%%%%%%%%%%%%%%%%

\begin{table*}
\centering
	\begin{tabular}{c c c c c c c c}
		\hline
		(1) & (2) & (3) & (4) & (5) & (6) & (7) & (8) \\\
		Model & Parameter & XMM1 & XMM2 & XMM3 & XMM4 & XMM5 & SUZ \\
		\hline
		\multicolumn{8}{c}{Partial Covering   ({\sc tbabs} $\times$ {\sc const} $\times$ {\sc zpcfabs} $\times$ {\sc zxipcf} $\times$ ({\sc zgauss} + {\sc powerlaw}))} \\
		\hline
		Constant & scale factor & $1^{f}$ & - & $0.96\pm0.02$ & - & - & - \\
		\hline
		Intrinsic & $\Gamma$ & $2.99\pm0.02$ & - & - & - & - & - \\
		Power Law & norm ($\times10^{-2}$) & $1.1\pm0.1$ & - & - & - & - & - \\
		\hline
		Neutral Absorber & nH ($\times10^{22}$ cm$^{-2}$) & $7\pm2$ & $3\pm1$ & $7^{l}$ & $4\pm1$ & $3.6\pm0.5$ & $7\pm2$ \\
		({\sc zpcfabs}) & CF & $0.43\pm0.07$ & $0.36\pm0.10$ & $0.43^{l}$ & $0.47\pm0.07$ & $0.50\pm0.02$ & $0.49\pm0.06$ \\
		\hline
		Ionised Absorber & nH ($\times10^{22}$ cm$^{-2}$) & $39\pm12$ & $19\pm6$ & $39^{l}$ & $43\pm9$ & $43\pm7$ & $38\pm11$ \\
		({\sc zxipcf}) & log($\xi$) (erg cm s$^{-1}$) & $0.4\pm0.5$ & - & - & - & - & - \\
		& CF & $0.62\pm0.06$ & $0.65\pm0.06$ & $0.62^{l}$ & $0.78\pm0.04$ & $0.75\pm0.03$ & $0.62\pm0.06$ \\
		\hline
		Ionised Iron & E (\kev) & $6.7\pm0.3$ & - & - & - & - & - \\
		Emission & $\sigma (\eV)$ & $1^{f}$ & - & - & - & - & - \\
		& Flux ($\times10^{-14}$ erg cm$^{-2}$ s$^{-1}$) & $1.4^{+1.5}_{-1.4p}$ & - & - & - & - & - \\
		\hline
		Flux & F$_{\rm 0.3-10}$ ($\times10^{-11}$ erg cm$^{-2}$ s$^{-1}$) & $1.23\pm0.02$ & $1.33\pm0.03$ & $1.18\pm0.02$ & $0.72\pm0.02$ & $0.76\pm0.01$ & $1.11\pm0.02$ \\
		\hline
		Fit Statistic & C/dof & 667/544 & - & - & - & - & - \\
		\hline
	\end{tabular}
	\caption{Best fit models for the partial covering scenario, including one neutral and one ionised partial covering component. The model component is given in column (1), and parameter names are in column (2). Parameters only quoted in the first column are kept linked between data sets. All parameters from XMM3 are linked to those of XMM1, denoted by the superscript "{\textit l}". All parameters with the superscript "{\textit f}" are kept fixed at quoted values. Parameters which have been pegged at a limit are denoted with "{\textit p}". The normalisation of the power law component is in units of photons keV$^{-1}$ cm$^{-2}$ s$^{-1}$ at 1keV.}
	\label{tab:covering}
\end{table*}

%%%%%%%%%%%%%%%%%%%%%%%%%%%%%%%%%%%%%%%%%%%%%%%%%%%%%%%%%%%%%%%%%%%%%%%%%%%%%%%%%%%%%%%%%%%%%%%%%

Another aspect of the absorption models is that they do not include the emission lines associated with the included absorption edges. The RGS data are consistent with the best-fitting model, but none of the expected emission lines are present in the RGS spectrum. Most notably, the $6.4\kev$ \feka\ emission line associated with the edge at $\sim7\kev$ is not included in the model. Assuming the obscuring sources are spherically symmetric, the predicted strength of the \feka\ line can be calculated by measuring the absorption strength of the edge (i.e., the drop in flux in the $7-20\kev$ range), and multiplying it by the fluorescent yield of iron (e.g. \citealt{Reynolds+2009}). If one assumes that only the neutral absorber is responsible for producing the $6.4\kev$ \feka\ feature, the resulting emission line is very weak, with an equivalent width of $4\eV$ using the absorption of the brightest spectrum (XMM2) and $5\eV$ using the dimmest spectrum (XMM4). No negative residuals are seen, suggesting that this line strength is consistent with a null detection. 

It is also, however, interesting to consider the ionisation on the other absorber. This value is in agreement with neutral, so this absorber would also contribute to the \feka\ line profile. This increases the equivalent width of the line, to $90\eV$ using the absorption of the brightest observations, and to $200\eV$ using the absorption of the dimmest. Both cases show significant negative residuals compared to the best-fitting model, implying that these lines should have been easily detectable in the observed spectrum. This suggests that non-spherically symmetric absorption is required to produce the observed spectra.

%%%%%%%%%%%%%%%%%%%%%%%%%%%%%%%%%%%%%%%%%%%%%%%%%%%%%%%%%%%%%%%%%%%%%%%%%%%%%%%%%%%%%%%%%%%%%%%%%
%%%%%%%%%%%%%%%%%%%%%%%%%%%%%%%%%%%%%%%%%%%%%%%%%%%%%%%%%%%%%%%%%%%%%%%%%%%%%%%%%%%%%%%%%%%%%%%%%

\subsection{Comptonisation}
\label{sect:optx}

The soft-Comptonisation (e.g. \citealt{Done+2012}) is characterised by a smooth soft excess as observed in the residuals in Fig.~\ref{fig:eeuf_plext}. This model was also suggested to explain the lack of features in the soft excess observed with the \chandra\ LETG detector by \cite{Marshall+2003}, and for the early \xmm\ observations by \cite{Guainazzi2004}.

To test this physical interpretation, the model {\sc optxagnf} (\citealt{Done+2012}) is used. This model describes the X-ray spectrum above $2\kev$ with a power law from a hot, spherical corona centred around the black hole. The soft X-ray spectrum is described with a second, cooler, optically thick corona located on top of the accretion disc. Emitted power is supplied by the energy released through accretion. The mass and accretion rate are fixed at values from \cite{Porquet+2004}, and the co-moving distance is set to 347 Mpc. Other free parameters include the radius of the primary, spherical corona (r$_{\rm cor}$), the temperature (kT) and opacity ($\tau$) of the soft X-ray corona, the slope of the hard Compton power law ($\Gamma$), and the fraction of power emitted in the hard Comptonisation component below r$_{\rm cor}$ (fpl). 

A variety of combinations of these parameters are tested to attempt to explain the variability between data sets. Allowing only the soft corona parameters (kT and $\tau$) or only the hard corona parameters ($\Gamma$ and fpl) to vary did not reproduce the spectral shape, with $C = 1045$ and $C = 1858$ for 554 dof respectively. Therefore, all four parameters are left free to vary between epochs. Allowing the radius of the primary corona to be free between data sets causes it to become unconstrained, so it is linked between epochs. Spin values ($a = cJ/GM^2$, where $M$ is the black hole mass and $J$ is the angular momentum) fixed at 0, 0.5 and 0.998 are tested, with a maximum spin giving the best fit. 

As with the partial covering model, excess residuals are visible in the $6-7\kev$ band, so a narrow ($\sigma = 1\eV$) Gaussian emission line is added. The best fit line energy and normalisation are free to vary, but are again linked between epochs. This improves the fit by $\Delta C = 30$ for 2 additional free parameters, for a final fit statistic of $C = 609$ for 546 dof. Again, no signatures of $6.4\kev$ emission are detected - the feature has a best fit energy of $6.6\pm0.1\kev$, consistent with the line energy found in the partial covering model. 

%%%%%%%%%%%%%%%%%%%%%%%%%%%%%%%%%%%%%%%%%%%%%%%%%%%%%%%%%%%%%%%%%%%%%%%%%%%%%%%%%%%%%%%%%%%%%%%%%

\begin{table*}
	\centering
	\begin{tabular}{c c c c c c c c}
		\hline
		(1) & (2) & (3) & (4) & (5) & (6) & (7) & (8) \\\
		Model & Parameter & XMM1 & XMM2 & XMM3 & XMM4 & XMM5 & SUZ \\
		\hline
		\multicolumn{8}{c}{{\sc optxagnf}   ({\sc tbabs} $\times$ {\sc const} $\times$ ({\sc zgauss} + {\sc optxagnf}))} \\
		\hline
		Constant & scale factor & $1^{f}$ & - & $0.96\pm0.02$ & - & - & - \\
		\hline
		Soft-Comptonisation & Mass ($\times10^{7}$ \Msun) & $1.99^{f}$ & - & - & - & - & - \\
		({\sc optxagnf}) & Distance (Mpc) & $347^{f}$ & - & - & - & - & - \\
		& log(L$_{\rm edd}$) & $-0.027^{f}$ & - & - & - & - & - \\
		& a & $0.998^{f}$ & - & - & - & - & - \\
		& r$_{\rm cor}$ (r$_{\rm g}$) & $64\pm5$ & - & - & - & - & - \\
		& log(r$_{\rm out}$) & $3^{f}$ & - & - & - & - & - \\
		& kT (\kev) & $0.26\pm0.03$ & $0.23\pm0.02$ & $0.26^{l}$ & $0.6\pm0.2$ & $0.57\pm0.12$ & $0.28\pm0.05$ \\
		& $\tau$ & $12.1\pm0.8$ & $13.2\pm0.8$ & $12.1^{l}$ & $7.2\pm1.2$ & $7.1\pm0.8$ & $11.5\pm1.3$ \\
		& $\Gamma$ & $2.23\pm0.05$ & $2.18\pm0.03$ & $2.23^{l}$ & $2.0\pm0.1$ & $2.01\pm0.06$ & $2.12\pm0.05$ \\
		& fpl & $0.13\pm0.01$ & $0.137\pm0.008$ & $0.13^{l}$ & $0.056\pm0.005$ & $0.061\pm0.003$ & $0.100\pm0.008$ \\
		\hline
		Ionised Iron & E (\kev) & $6.6\pm0.1$ & - & - & - & - & - \\
		Emission & $\sigma (\eV)$ & $1^{f}$ & - & - & - & - & - \\
		& Flux & $1.5\pm0.8$ & - & - & - & - & - \\
		&($\times10^{-14}$ erg cm$^{-2}$ s$^{-1}$) & & & & & & \\
		\hline
		Flux & F$_{\rm 0.3-10}$ & $1.23\pm0.04$ & $1.33\pm0.03$ & $1.18\pm0.03$ & $0.7\pm0.3$ & $0.8\pm0.3$ & $\sim1.12$ \\
		& ($\times10^{-11}$ erg cm$^{-2}$ s$^{-1}$) & & & & & & \\
		\hline
		Fit Statistic & C/dof & 609/546 & - & - & - & - & - \\
		\hline
	\end{tabular}
	\caption{Best fit models for the soft-Comptonisation model {\sc optxagnf}. The model component is given in column (1), and parameter names are in column (2). Parameters only quoted in the first column are kept linked between data sets. All parameters from XMM3 are linked to those of XMM1, denoted by the superscript "{\textit l}". All parameters with the superscript "{\textit f}" are kept fixed at quoted values.}
	\label{tab:optxagnf}
\end{table*}

%%%%%%%%%%%%%%%%%%%%%%%%%%%%%%%%%%%%%%%%%%%%%%%%%%%%%%%%%%%%%%%%%%%%%%%%%%%%%%%%%%%%%%%%%%%%%%%%%

\begin{figure}
	\includegraphics[width=\columnwidth]{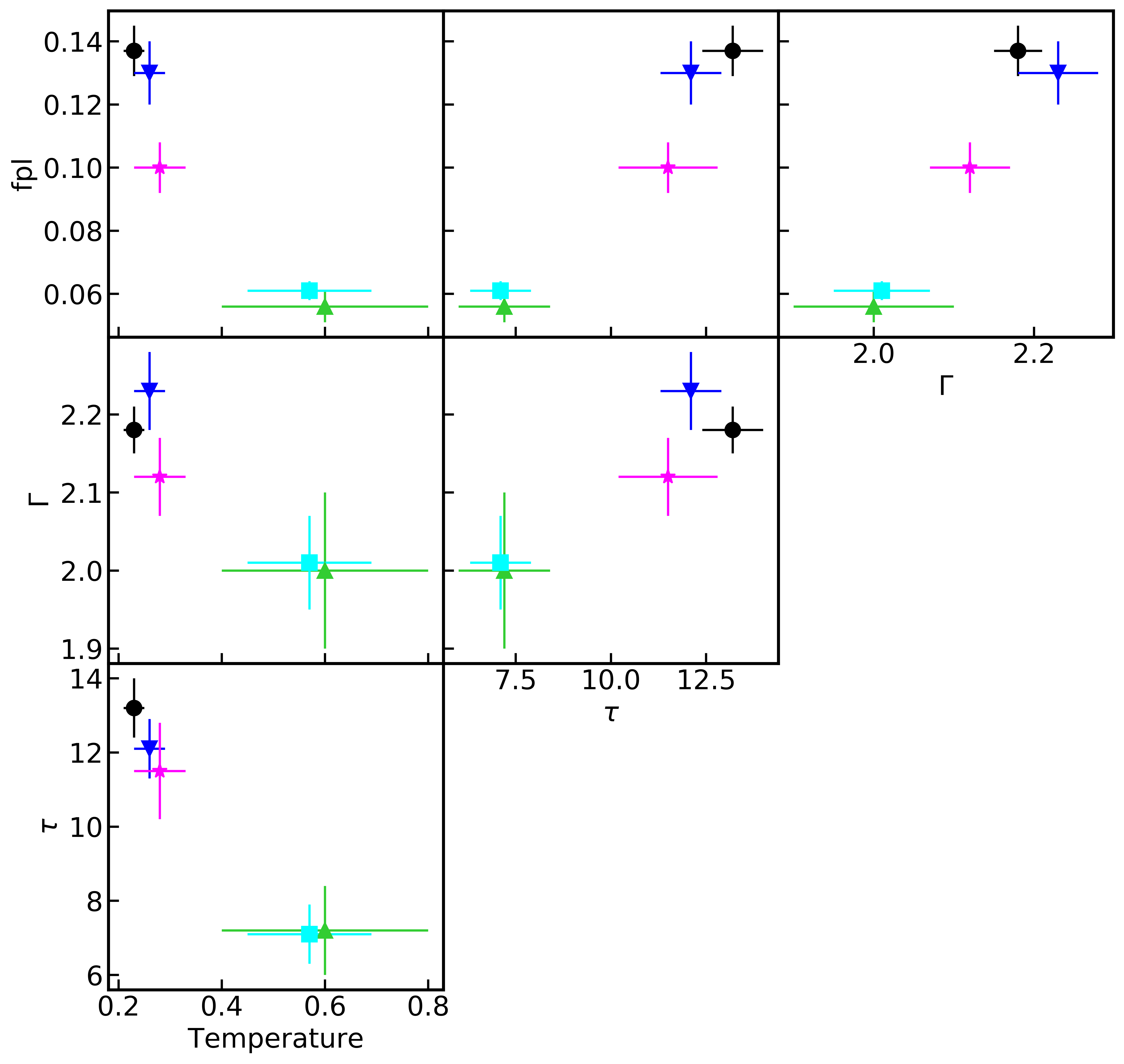}
	\caption{Correlations between best fit parameters using the Comptonisation model. Clear correlations are present between all parameters, and changes between brighter and dimmer flux epochs are evident in all parameters.}
	\label{fig:optxcorr}
\end{figure}

The best fit model is shown in the top right panel of Fig.~\ref{fig:model}, where the effects of Galactic absorption have been removed for display. Best fit parameters are listed in Table~\ref{tab:optxagnf}, and correlations between parameters are shown in Fig.~\ref{fig:optxcorr}. All parameters appear highly correlated with one another, and all change according to the flux of each spectrum. For the dimmer XMM4 and XMM5, the temperature of the secondary corona increases, while the opacity, slope of the hard power law, and fpl all decrease compared to the brighter data sets. All parameters are well constrained and vary significantly between the brighter and dimmer flux epochs. Despite the AGN being at a bright flux during the SUZ epoch, the parameter values are more intermediate to the high and low states. This could be due to the more limited band pass studied with \suzaku.

%%%%%%%%%%%%%%%%%%%%%%%%%%%%%%%%%%%%%%%%%%%%%%%%%%%%%%%%%%%%%%%%%%%%%%%%%%%%%%%%%%%%%%%%%%%%%%%%%

%%%%%%%%%%%%%%%%%%%%%%%%%%%%%%%%%%%%%%%%%%%%%%%%%%%%%%%%%%%%%%%%%%%%%%%%%%%%%%%%%%%%%%%%%%%%%%%%%

Another interesting test of the {\sc optxagnf} model is to take the best fit model to the X-ray data and apply it to optical/UV data, as this model is intended for broad SED fitting. To do so, the tool {\sc ftflx2xsp} is used to build dummy response files suitable for use in {\sc xspec} for the available UVW1 and UVW2 data. Only these filters are used to avoid potential host galaxy contamination. We then extrapolate the best-fitting models for both a maximum spin ($a=0.998$) and non-spinning ($a=0$) black hole to the UV data and examine the fit. 

%%%%%%%%%%%%%%%%%%%%%%%%%%%%%%%%%%%%%%%%%%%%%%%%%%%%%%%%%%%%%%%%%%%%%%%%%%%%%%%%%%%%%%%%%%%%%%%%%

\begin{figure}
	\includegraphics[width=\columnwidth]{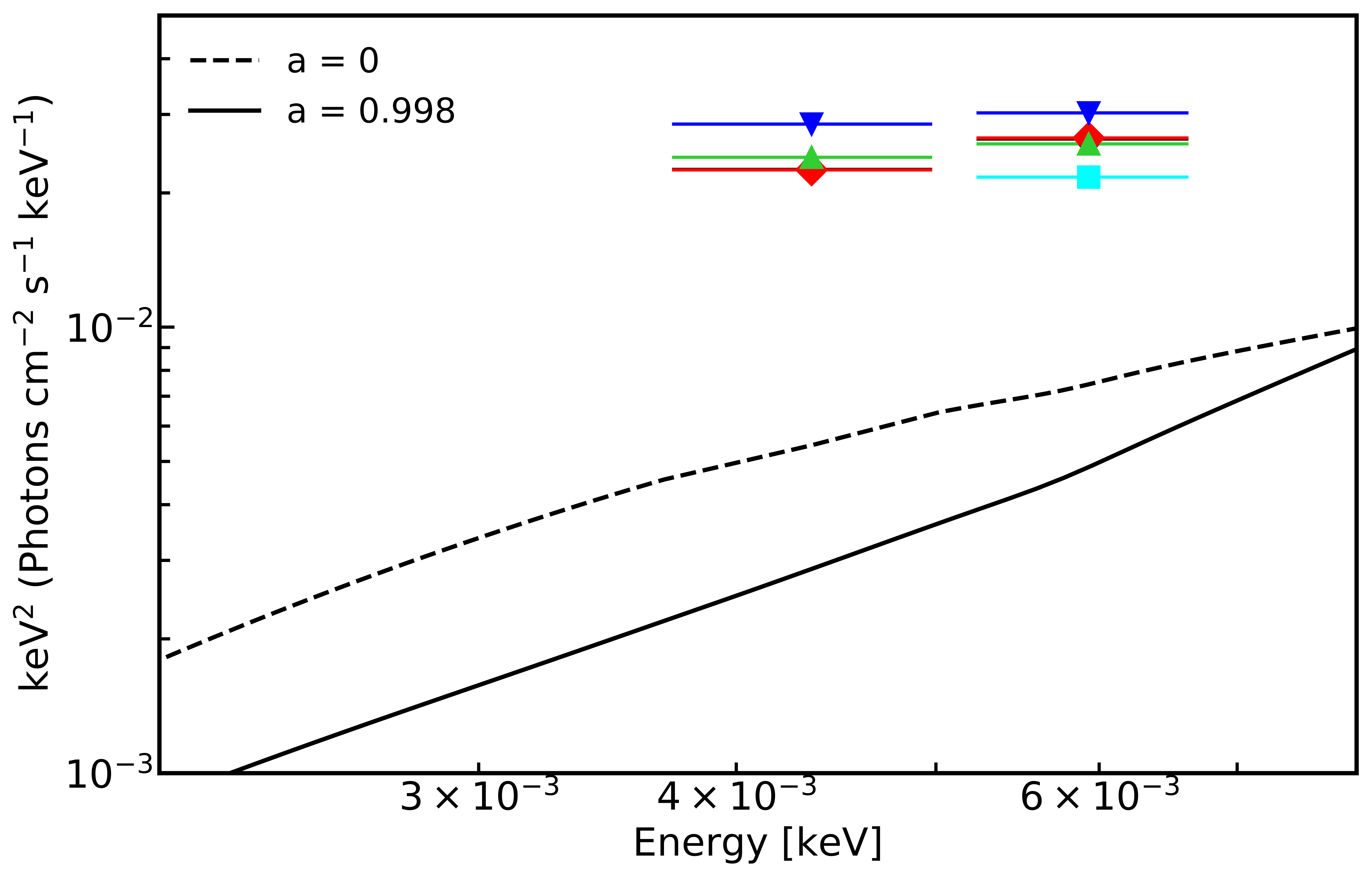}
	\caption{All available UVW1 and UVW2 data for all epochs compared to the a = 0 (dashed) and a = 0.998 (solid) best fit Comptonisation models. Colours and shapes match those of the corresponding X-ray data sets. Y-axis error bars on the UV data are shown, but are smaller than the symbols. The best fit models to the X-ray data underestimate the data when extrapolated to UV energies.}
	\label{fig:optxuv}
\end{figure}

%%%%%%%%%%%%%%%%%%%%%%%%%%%%%%%%%%%%%%%%%%%%%%%%%%%%%%%%%%%%%%%%%%%%%%%%%%%%%%%%%%%%%%%%%%%%%%%%%

The result is shown in Fig.~\ref{fig:optxuv}. The colours and symbols used on the optical data match those of the corresponding X-ray spectra. No simultaneous optical data is available for the \suzaku\ data. The best fit models at spins of 0.998 and 0 are shown as black solid and dashed lines, respectively. Models for all data sets are the same at this low energy range, so only one model line is shown for each spin value. Neither model is able to reproduce the shape or flux of the data at these low energies. Both underestimate the UV flux, by a factor of $\sim5$ for the UVW2 data and $\sim10$ for UVW1, assuming maximum spin, and a more modest factor of $\sim3.5$ for UVW2 and $\sim5$ for UVW1 assuming no spin. Allowing the Eddington luminosity to go free results in a best-fitting super-Eddington accretion rate (log(L/L$_{\rm edd}$)$ \simeq0.14$) and does not improve the fit to the UV data. Similarly, simultaneous modelling of the UV and X-ray data fails to find a fit which explains the UV flux. This shows that the soft-Comptonisation model is unable to explain the observed SED for Mrk 478.

%%%%%%%%%%%%%%%%%%%%%%%%%%%%%%%%%%%%%%%%%%%%%%%%%%%%%%%%%%%%%%%%%%%%%%%%%%%%%%%%%%%%%%%%%%%%%%%%%
%%%%%%%%%%%%%%%%%%%%%%%%%%%%%%%%%%%%%%%%%%%%%%%%%%%%%%%%%%%%%%%%%%%%%%%%%%%%%%%%%%%%%%%%%%%%%%%%%

\subsection{Blurred Reflection}
\label{sect:refl}

In the blurred reflection model, the intrinsic power law is seen alongside a reflection spectrum, produced when X-ray photons from the corona strike the inner accretion disc. This model has been used successfully to explain the spectral properties and variability of numerous NLS1 galaxies (e.g. \citealt{Fabian+2004,Ponti+2010,Gallo+2019m335}). It has also been used to explain time domain variability and lags (e.g. \citealt{Wilkinslags}). This interpretation was also discussed in detail by \cite{Zoghbi+2008}, who found that a highly blurred, highly ionised, reflection dominated model explained the spectra of XMM1 - XMM4.   

To test the blurred reflection interpretation, two different models are used; {\sc relxill} version 1.2.0 (\citealt{relxillmodel}) and {\sc reflionx} (\citealt{Ross+1999,RossFabian+2005}). {\sc reflionx} was convolved by the blurring model {\sc kerrconv} (\citealt{kerrmodel}). {\sc reflionx} was combined with a more simplistic blurring model {\sc kdblur} for analysis in \cite{Zoghbi+2008}. Both of these models are combined with a power law to model the intrinsic coronal emission. 

To measure the reflection fraction, the convolution model {\sc cflux} was applied to both the reflection and power law components. The flux parameter was linked between the two models. The relative fluxes of the models were then determined by adding a constant between the reflection and power law components. This constant measures the reflection fraction, R, the ratio of emitted flux in the reflection and power law components in the $0.1-100\kev$ band.

Parameters are the same between models. Both include two emissivity index values q$_{\rm in}$ and q$_{\rm out}$, separated at a break radius r$_{\rm br}$. These parameters define the illumination pattern, which goes as $\epsilon \propto r^{-q}$. The inner emissivity index is free to vary between epochs, while the break radius and outer emissivity index are kept fixed at 6 r$_g$ and 3, respectively. The inner radius of the accretion disc is kept fixed at the innermost stable circular orbit (ISCO), while the outer radius is fixed at 400 r$_{g}$, as little emission is expected to originate from outside this radius. The spin ($a$), inclination and iron abundance are left free to vary, but linked between epochs, as none are expected to vary within the given timescales. The {\sc xspec} function {\sc steppar} is also run on these parameters to ensure that none are confined to local minima. 

The flux produced by the power law component is left free to vary between epochs, as is the photon index ($\Gamma$). The photon index of the reflection component is linked to that of the power law, as leaving it free to vary does not improve the fit. Various combinations between other parameters are tested, and it is found that the best fit is produced when the reflection fraction and ionisation (again defined as $\xi = 4\pi F/n$) are left free to vary between observations, although the changes in these parameters are limited within uncertainties between epochs.

As in the other models, a narrow feature is added to fit the residuals apparent between $6-7\kev$. The best fit line energies are identical for both models, at $6.6\pm0.1\kev$. Once again, no evidence for narrow $6.4\kev$ emission is detected. As such, no distant reflection model to account for reflection off of the neutral torus is added to the model.

The best fit parameters are shown in Table~\ref{tab:reflection}, and the models and residuals are shown in the bottom two panels of Fig.~\ref{fig:model}. Both models produced comparable fits and have the same degrees of freedom (539).  The C-statistic is $593$ and $606$ for the {\sc relxill} and {\sc reflionx} models, respectively.  However, each model gives a different interpretation of the X-ray emitting region.

%%%%%%%%%%%%%%%%%%%%%%%%%%%%%%%%%%%%%%%%%%%%%%%%%%%%%%%%%%%%%%%%%%%%%%%%%%%%%%%%%%%%%%%%%%%%%%%%%

\begin{table*}
\centering
	\begin{tabular}{c c c c c c c c}
		\hline
		(1) & (2) & (3) & (4) & (5) & (6) & (7) & (8) \\\
		Model & Parameter & XMM1 & XMM2 & XMM3 & XMM4 & XMM5 & SUZ \\
		\hline
		\multicolumn{8}{c}{{\sc reflionx}   ({\sc tbabs} $\times$ {\sc const$_1$} $\times$ ({\sc zgauss} + ({\sc cflux} $\times$ {\sc cutoffpl}) + ({\sc const$_2$} $\times$ {\sc cflux} $\times$ {\sc kerrconv} $\times$ {\sc reflionx}))} \\
		\hline
		Constant & scale factor & $1^{f}$ & - & $0.89\pm0.06$ & - & - & - \\
		\hline
		Power Law & $\Gamma$ & $2.73\pm0.06$ & $2.7\pm0.1$ & $2.73^{l}$ & $2.6\pm0.1$ & $2.65\pm0.06$ & $2.8\pm0.1$ \\
		& E$_{\rm cut} (\kev)$ & $300^{f}$ & - & - & - & - & - \\
		& log(F$_{0.1-100}$) & $-10.67\pm0.05$ & $-10.66\pm0.09$ & $-10.67^{l}$ & $-11.0\pm0.1$ & $-10.91\pm0.04$ & $-10.8\pm0.1$ \\
		\hline
		Blurring & R$_{\rm in}$ (ISCO) & 1$^{f}$ & - & - & - & - & - \\
		({\sc kerrconv}) & R$_{out}$ (r$_{g}$) & 400$^{f}$ & - & - & - & - & - \\
		& R$_{br}$ (r$_{g}$) & 6$^{f}$ & - & - & - & - & - \\
		& q$_{\rm in}$ & $8\pm1$ & $9.0^{+0.8}_{-1.4}$ & $8^{l}$ & $9.2^{+0.8p}_{-1.4}$ & $8.9\pm1.0$ & $9.4^{+0.6p}_{-1.2}$\\
		& q$_{\rm out}$ & 3$^{f}$ & - & - & - & - & - \\
		& a & $0.94\pm0.02$ & - & - & - & - & - \\
		& i (\degree) & $<22$ & - & - & - & - & - \\
		\hline
		Reflection & $\xi$ (erg cm s$^{-1}$) & $63\pm35$ & $55\pm40$ & $63^{l}$ & $<165$ & $40\pm22$ & $85\pm63$ \\
		({\sc reflionx}) & A$_{\rm Fe}$ (Fe/solar) & $0.44\pm0.28$ & - & - & - & - & - \\
		& R$_{0.1-100}$ & $0.7\pm0.3$ & $0.8\pm0.3$ & $0.7^{l}$ & $1.0\pm0.5$ & $0.7\pm0.1$ & $1.3\pm0.6$ \\
		\hline
		Ionised Iron & E (\kev) & $6.6\pm0.1$ & - & - & - & - & - \\
		Emission & $\sigma (\kev)$ & $0.001^{f}$ & - & - & - & - & - \\
		& Flux ($\times10^{-14}$ erg cm$^{-2}$ s$^{-1}$) & $1.3\pm0.8$ & - & - & - & - & - \\
		\hline
		Flux & F$_{\rm 0.3-10}$ ($\times10^{-11}$ erg cm$^{-2}$ s$^{-1}$) & $1.23\pm0.02$ & $1.35\pm0.03$ & $1.18\pm0.03$ & $0.7\pm0.1$ & $0.76\pm0.02$ & $1.2\pm0.3$ \\
		\hline
		Fit Statistic & C/dof & 606/539 & - & - & - & - & - \\
		\hline
		\hline
		\multicolumn{8}{c}{{\sc relxill}   ({\sc tbabs} $\times$ {\sc const$_1$} $\times$ ({\sc zgauss} + ({\sc cflux} $\times$ {\sc cutoffpl}) + ({\sc const$_2$} $\times$ {\sc cflux} $\times$ {\sc relxill}))} \\
		\hline
		Constant & scale factor & $1^{f}$ & - & $0.96\pm0.02$ & - & - & - \\
		\hline
		Power Law & $\Gamma$ & $2.51\pm0.04$ & $2.47\pm0.06$ & $2.51^{l}$ & $2.5\pm0.1$ & $2.42\pm0.03$ & $2.6\pm0.2$ \\
		& E$_{\rm cut} (\kev)$ & $300^{f}$ & - & - & - & - & - \\
		& log(F$_{0.1-100}$) & $-11.1\pm0.2$ & $-11.0\pm0.1$ & $-11.1^{l}$ & $-11.2\pm0.1$ & $-11.25\pm0.08$ & $-11.0\pm0.2$ \\
		\hline
		Blurred & R$_{\rm in}$ (ISCO) & $1^{f}$ & - & - & - & - & - \\
		Reflection & R$_{\rm out}$ (r$_{g}$) & $400^{f}$ & - & - & - & - & - \\
		({\sc relxill}) & R$_{br}$ (r$_{g}$) & $6^{f}$ & - & - & - & - & - \\
		& q$_{\rm in}$ & $6.9\pm0.6$ & $7.3\pm0.8$ & $6.9^{l}$ & $8.0\pm1.6$ & $7.8\pm0.7$ & $8.7^{+1.3p}_{-1.8}$ \\
		& q$_{\rm out}$ & $3^{f}$ & - & - & - & - & - \\
		& a & $0.98\pm0.01$ & - & - & - & - & - \\
		& i (\degree) & $31\pm8$ & - & - & - & - & - \\
		& log($\xi$) (erg cm s$^{-1}$) & $3.1\pm0.1$ & $2.8\pm0.2$ & $3.1^{l}$ & $2.8\pm0.3$ & $2.9\pm0.1$ & $2.7\pm0.4$ \\
		& A$_{\rm Fe}$ (Fe/solar) & $0.84\pm0.14$ & - & - & - & - & - \\
		& R$_{0.1-100}$ & $3.5\pm1.5$ & $2.3\pm0.9$ & $3.5^{l}$ & $2.0\pm0.9$ & $2.4\pm0.6$ & $3.1\pm1.7$ \\
		\hline
		Ionised Iron & E \kev & $6.6\pm0.1$ & - & - & - & - & - \\
		Emission & $\sigma \kev$ & $0.001^{f}$ & - & - & - & - & - \\
		& Flux ($\times10^{-14}$ erg cm$^{-2}$ s$^{-1}$) & $1.5\pm0.8$ & - & - & - & - & - \\
		\hline
		Flux & F$_{\rm 0.3-10}$ ($\times10^{-11}$ erg cm$^{-2}$ s$^{-1}$) & $1.2\pm0.4$ & $1.33\pm0.06$ & $1.2\pm0.4$ & $0.71\pm0.04$ & $0.8\pm0.2$ & $1.2\pm0.1$ \\
		\hline
		Fit Statistic & C/dof & 593/539 & - & - & - & - & - \\
		\hline
	\end{tabular}
	\caption{Best fit models for the blurred reflection models {\sc reflionx} (\textbf{top}) and {\sc relxill} (\textbf{bottom}). Note that const$_1$ refers to the scaling factor between XMM1 and XMM3, while const$_2$ is used to measure the reflection fraction, R. The model component is given in column (1), and parameter names are in column (2). Parameters only quoted in the first column are kept linked between data sets. All parameters from XMM3 are linked to those of XMM1, denoted by the superscript "{\textit l}". All parameters with the superscript "{\textit f}" are kept fixed at quoted values. Parameters which have been pegged at a limit are denoted with "{\textit p}".}
	\label{tab:reflection}
\end{table*}

%%%%%%%%%%%%%%%%%%%%%%%%%%%%%%%%%%%%%%%%%%%%%%%%%%%%%%%%%%%%%%%%%%%%%%%%%%%%%%%%%%%%%%%%%%%%%%%%%

\begin{figure*}
	\begin{minipage}{0.48\textwidth}
		\includegraphics[width=\columnwidth]{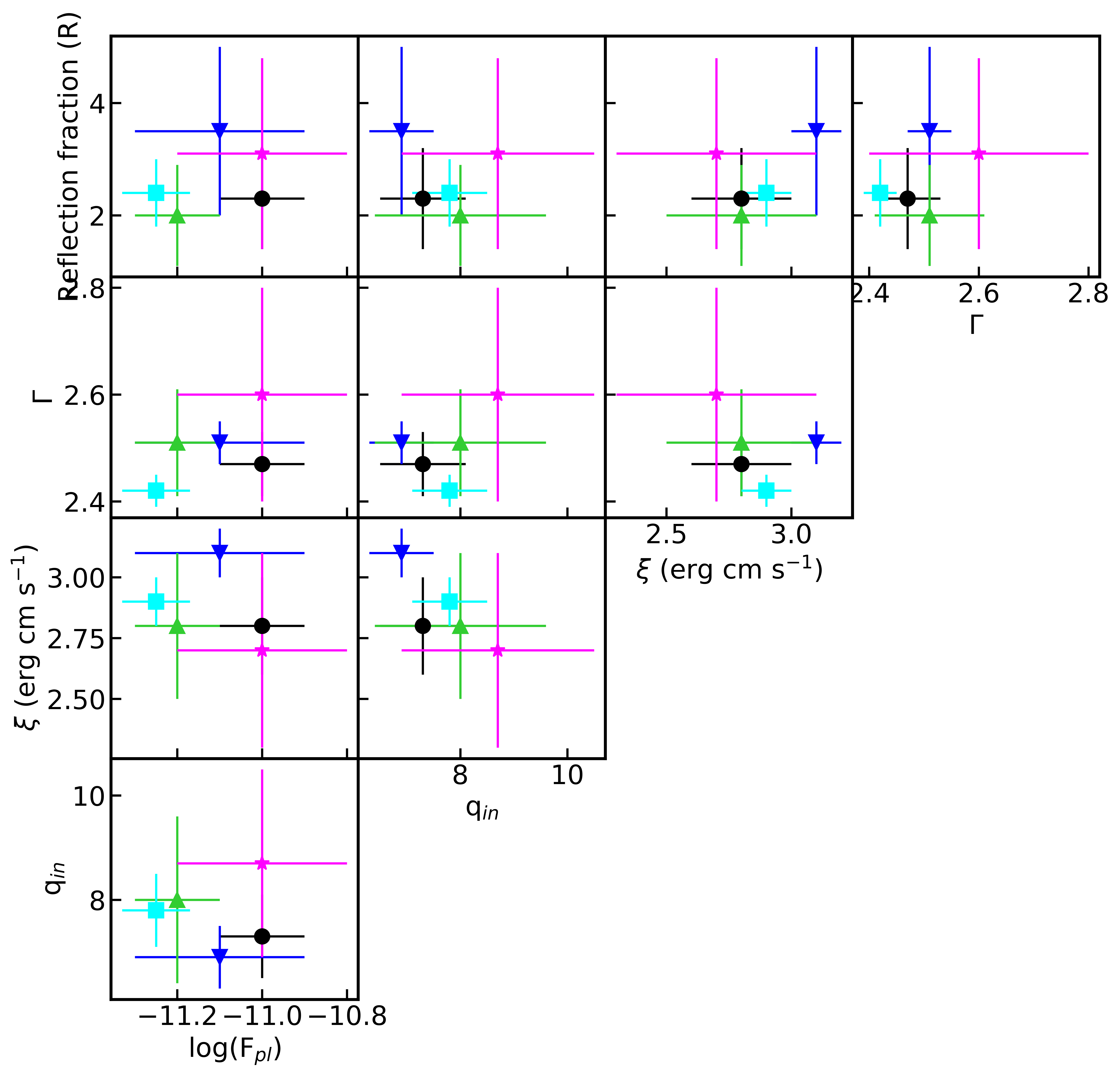}
	\end{minipage}
	\begin{minipage}{0.48\textwidth}
		\includegraphics[width=\columnwidth]{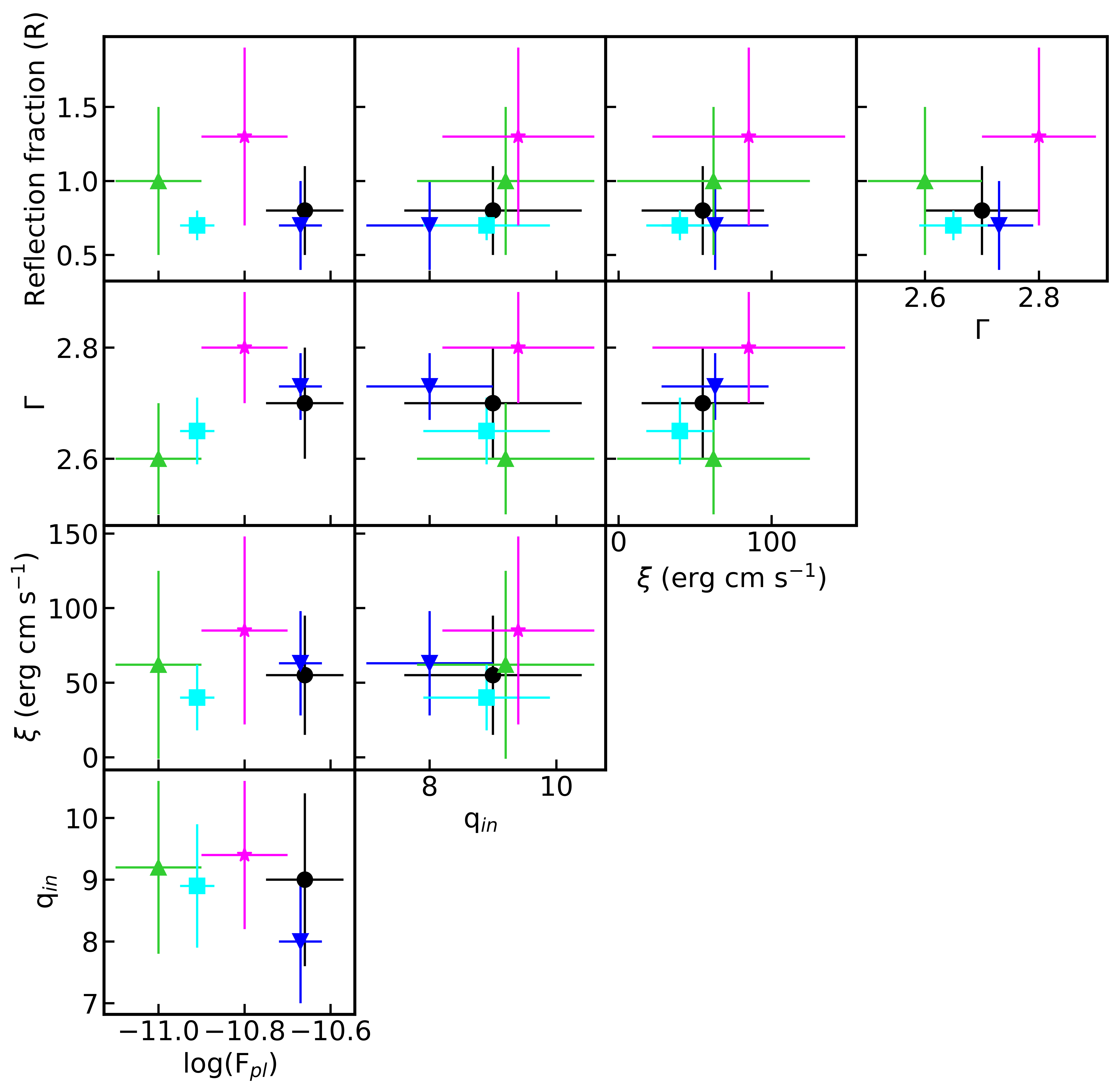}
	\end{minipage}
	\caption{Correlations between best fit parameters using {\sc relxill} (left) and {\sc reflionx} (right). Although the error bars are large, {\sc reflionx} shows more evidence for correlation between F$_{\rm pl}$ and $\Gamma$ with $\xi$, and $\xi$ and q$_{\rm in}$ with R. Very little evidence for any correlations can be seen using {\sc relxill}, between any parameters.}
	\label{fig:refcorr}
\end{figure*}

To begin comparing the two models, correlation plots are made using the best fitting free parameters. The results are presented in Fig.~\ref{fig:refcorr}, with {\sc relxill} parameters shown on the left and {\sc reflionx} parameters shown on the right. Axes between figures are not identical, but corresponding cells in each panel show correlations between the same parameters. The colours and styles of the data points match those used throughout to represent data sets. Points for XMM3 are not shown, as these parameters have all been linked to XMM1 and are not independently measured. 

The inner emissivity is in agreement, although poorly constrained, between both models for all data sets. For the power law component, photon indices are generally steeper using {\sc reflionx} than for {\sc relxill}, and do not agree within error. This is also true for the power law fluxes, which are generally higher (brighter) for {\sc reflionx}. For XMM4 and SUZ, however, both of these values agree between models, as these data sets most poorly constrain the parameters. None of the spectra are as steep as the value measured using partial covering, and are significantly steeper than when using soft-Comptonisation.

\begin{figure}
	\includegraphics[width=\columnwidth]{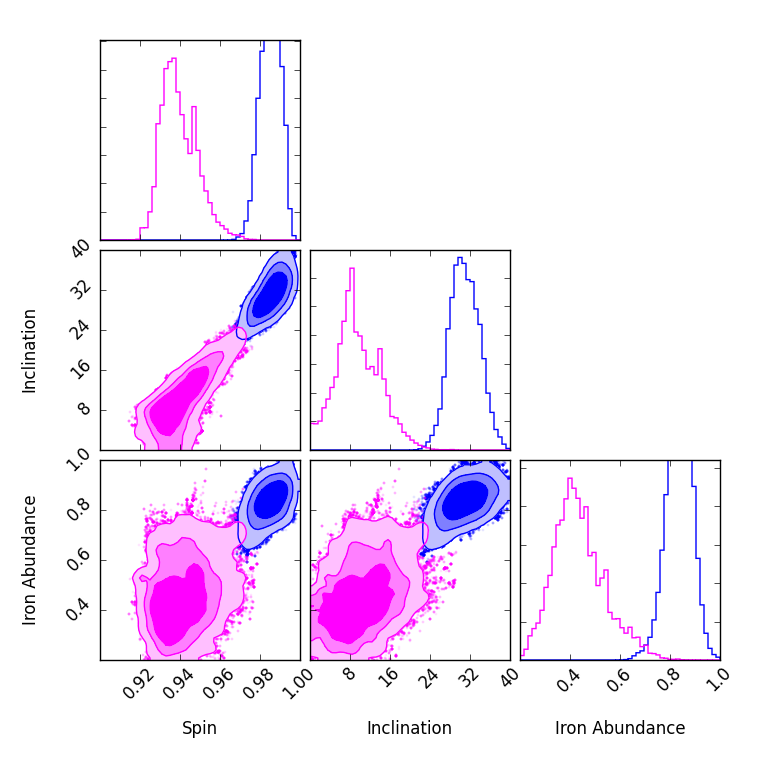}
	\caption{Comparison between the best fit spin, inclination and iron abundance using {\sc relxill} (blue) and {\sc reflionx} (pink). Contours are shaded using 68, 90 and 99 percent of MCMC fits.}
	\label{fig:relvsref_corner}
\end{figure}

Despite the fact that not all variable parameters agree between the two reflection models represented here, the parameters kept constant between data sets are in agreement with one another; notably black hole spin ($a$), inclination and iron abundance. This is demonstrated in Fig.~\ref{fig:relvsref_corner}. Data for {\sc relxill} are shown in blue and data from {\sc reflionx} are in pink. Contours are produced using 68, 90 and 99 percent of MCMC fits. The histograms reveal that all parameters are evenly distributed around the measured mean. The contours also reveal that all parameters are in agreement, and confirm that the reflection interpretation requires a high spin, low inclination AGN. 

It is in the interpretation of the reflection fraction and ionisation parameters that the two reflection models are mostly extremely in disagreement. {\sc relxill} suggests a highly ionised spectrum, with $\xi\simeq1000\ergpscmps$, while {\sc reflionx} predicts a lower value of $\xi\simeq50\ergpscmps$. Setting the ionisation parameters of {\sc relxill} to the ones found by {\sc reflionx} and vice versa degrade the fit quality by $\Delta C = 146$ and $\Delta C = 40$, respectively. For the reflection fraction, use of {\sc relxill} suggests a reflection dominated spectrum, with all reflection fractions band constrained to be larger than one, and the largest being less than 3.5. However, using {\sc reflionx} suggests a power law dominated spectrum, or one with equal contributions from the intrinsic power law and reflection spectra, with R values constrained between $\sim0.7-1.3$ for all spectra. 

As seen in Fig.~\ref{fig:refcorr}, using {\sc relxill} suggests that parameters for all data sets are mostly in agreement within error, and no correlations between parameters are apparent. This is not the case with {\sc reflionx}. There is a clear positive correlation between the photon index $\Gamma$ and the flux of the power law component. This relationship is expected; as $\Gamma$ increases, more emission is expected in the soft band, increasing the overall flux. Interestingly, it is also for {\sc reflionx} that we see clear changes in flux state; with {\sc relxill}, the flux between epochs does not change within error. For {\sc reflionx}, weaker evidence for correlations are also visible between the reflection fraction, inner emissivity index and ionisation, but these parameters are not well constrained.

The two reflection models presented here ({\sc relxill} and {\sc reflionx}) are both dominated by a single spectral component in the $0.3-10\kev$ band. For {\sc relxill}, this is the reflection spectrum, while {\sc reflionx} suggests the power law dominates. Changes in spectra are caused primarily by flux variations of these components. Both interpretations seem consistent with the approximately constant hardness ratio found in Section~\ref{sect:var} (see also Section~\ref{sect:pca}).

%%%%%%%%%%%%%%%%%%%%%%%%%%%%%%%%%%%%%%%%%%%%%%%%%%%%%%%%%%%%%%%%%%%%%%%%%%%%%%%%%%%%%%%%%%%%%%%%%%

%%%%%%%%%%%%%%%%%%%%%%%%%%%%%%%%%%%%%%%%%%%%%%%%%%%%%%%%%%%%%%%%%%%%%%%%%%%%%%%%%%%%%%%%%%%%%%%%%

%%%%%%%%%%%%%%%%%%%%%%%%%%%%%%%%%%%%%%%%%%%%%%%%%%%%%%%%%%%%%%%%%%%%%%%%%%%%%%%%%%%%%%%%%%%%%%%%%
%%%%%%%%%%%%%%%%%%%%%%%%%%%%%%%%%%%%%%%%%%%%%%%%%%%%%%%%%%%%%%%%%%%%%%%%%%%%%%%%%%%%%%%%%%%%%%%%%

\subsection{Principal Component Analysis}
\label{sect:pca}

To obtain a model-independent assessment of the spectral variations between observations, principal component analysis (PCA) is used. This technique is implemented using the method and code{\footnote{http://www-xray.ast.cam.ac.uk/$\sim$mlparker/}} described by \cite{Parker+2014}. By finding the eigenvalues corresponding to the maximum spectral variability, it allows for the detection of correlated variability between energy bands.

\begin{figure}
	\includegraphics[width=\columnwidth]{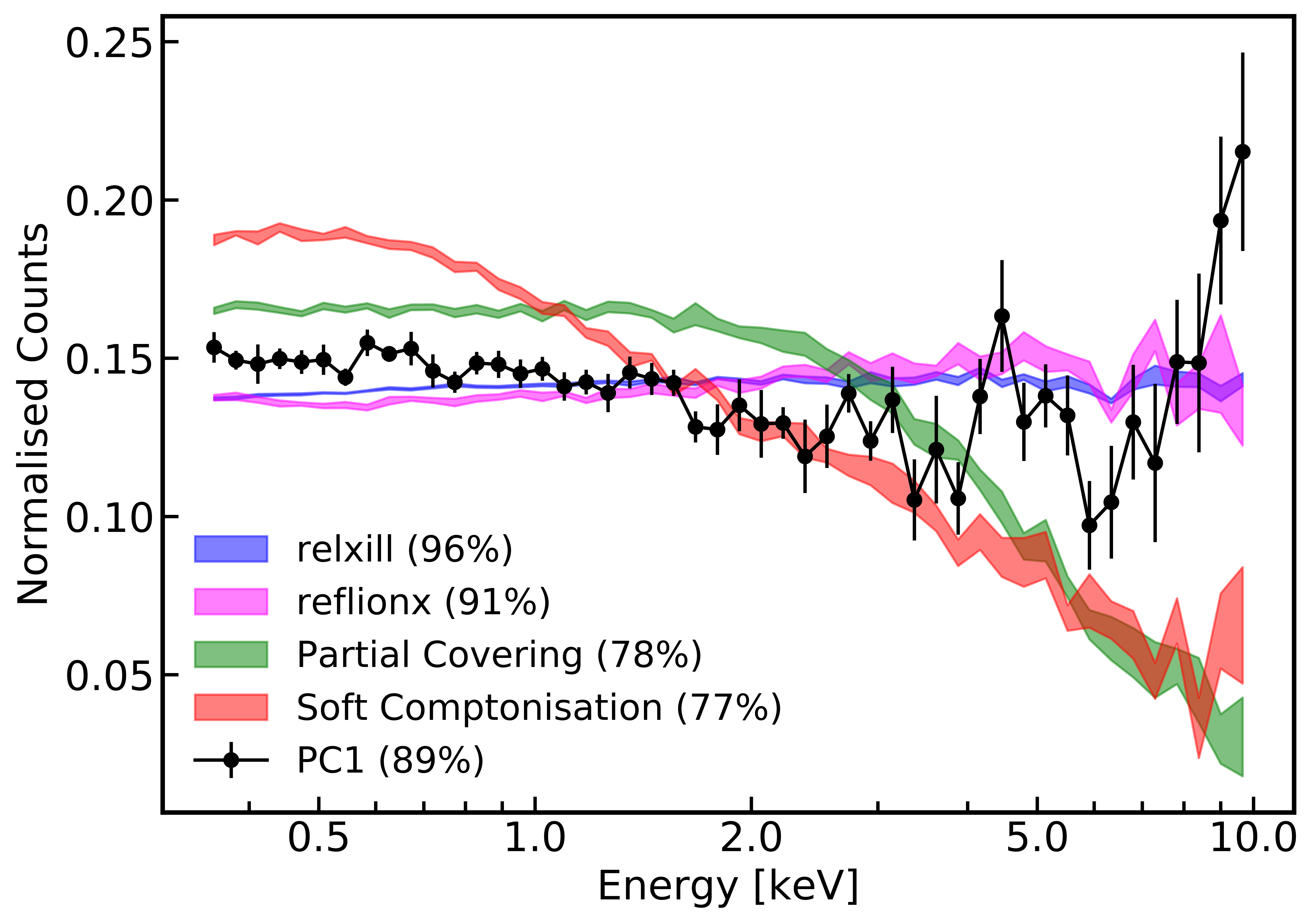}
	\caption{PCA using 20ks segments for all \xmm\ observations (black). Only PC1 is significant, and accounts for $90$ per cent of the variability. The shape is relatively flat, with some curvature around $4-7\kev$. The PCA results from simulated data sets are also shown, with partial covering in green, {\sc optxagnf} in red, {\sc relxill} in blue and {\sc reflionx} in pink. The blurred reflection models best reproduce the flat shape of PC1.}
	\label{fig:pca}
\end{figure}

To investigate long-term variability trends, data from all \xmm\ observations are combined and the full $0.3-10\kev$ band is used. The long exposure (XMM5) is broken into seven $20\ks$ segments so all spectra have similar exposures. Only one significant principal component (PC1) is revealed in this analysis, accounting for $\sim90$ per cent of the variability. This component is shown as a function of energy in Fig.~\ref{fig:pca}. The shape is mostly flat, with some curvature in the $4-7\kev$ region. There is also an upturn above $\sim8.5\kev$, which may be  attributable to variations in the background spectrum between epochs. All PC values are positive, indicating that the changes in each energy band are correlated. 

\cite{Parkerpca} and \cite{Gallant+2018} are able to reproduce this shape with a single variable model component (e.g. a power law) that varies in normalisation. This implies that the spectral changes between epochs can mostly be attributed to flux changes of a single model component, such as the power law. This interpretation is consistent with the minimal changes in hardness ratio within and between observations, and lack of suggested spectral state change based on $\Delta \alpha_{\rm ox}$ (see Section~\ref{sect:var}). Significant changes in spectral parameters in Mrk 478 are not expected based on the PCA. Such changes typically yield more complex principal component shapes, as outlined in \cite{Parkerpca}. It is important to note that more complex physical scenarios, such as multiple variable absorption zones, can in some cases produce an overall flat PCA shape (see \citealt{Miller+2008}), however, the available data are insufficient to model with more complex scenarios.

PCA also provides a unique way to assess the capability of the models above to reproduce the observed variability. To do so, 100 fake data sets with $20\ks$ exposures are simulated for each model. Parameters which are linked between data sets are frozen for all simulations. Free parameters are varied between each simulated data set, with ranges based on model constraints. Correlations between parameters are not apparent using partial covering or {\sc relxill} and very weak using {\sc reflionx}, but all parameters are highly correlated for {\sc optxagnf}. A PCA is then produced for these simulated data sets.

The results are shown in Fig.~\ref{fig:pca}. The width of the bands represent the calculated error bars from the PCA analysis. All models predict correlated variability between all energy bands, but differ in shape of the first principal component. The PCA results for partial covering and {\sc optxagnf} exhibit significant curvature, turning sharply downwards towards high energies. Both reflection models, however, produce first principle components which are fairly constant across all energies. This is likely due to the fact that most variations in these models are in the flux of the dominant emission component. For {\sc relxill}, this corresponds to flux changes in the reflection spectrum, whereas for {\sc reflionx}, this is the power law component. This is consistent with the simulations presented in \cite{Parkerpca} and \cite{Gallant+2018}, where normalisation changes of a single spectral component produced the PCA shapes.

To evaluate the statistical fit of each model, $\redchi$ values are obtained for each simulated PCA compared to the data. For 50 degrees of freedom, the $\redchi$ values are 17, 39, 4 and 6 for partial covering, soft-Comptonisation, {\sc relxill} and {\sc reflionx}, respectively. The measured statistics, as well as the overall flatter shapes produced by the blurred reflection models, suggest that this interpretation best fits the observed PCA. Although the varied parameters cannot exactly reproduce the observed variability, the overall shape is very close. The other models are clearly unable to reproduce the almost flat shape of the PCA calculated using all \xmm\ observations.

%%%%%%%%%%%%%%%%%%%%%%%%%%%%%%%%%%%%%%%%%%%%%%%%%%%%%%%%%%%%%%%%%%%%%%%%%%%%%%%%%%%%%%%%%%%%%%%%%
%%%%%%%%%%%%%%%%%%%%%%%%%%%%%%%%%%%%%%%%%%%%%%%%%%%%%%%%%%%%%%%%%%%%%%%%%%%%%%%%%%%%%%%%%%%%%%%%%
%%%%%%%%%%%%%%%%%%%%%%%%%%%%%%%%%%%%%%%%%%%%%%%%%%%%%%%%%%%%%%%%%%%%%%%%%%%%%%%%%%%%%%%%%%%%%%%%%

\section{Discussion}
\label{sect:discussion}

\subsection{The X-ray nature of Mrk 478}
\label{sect:discussionxray}

The spectra of Mrk 478 at all epochs are similar, characterised by a strong, smooth soft excess and excess residuals in the $5-7\kev$ band, attributable to some form of iron emission. The variation between epochs is primarily due to normalisation changes, and little evidence for shape changes are found. The data also do not require any additional emission or absorption features produced in a warm absorber. This is supported by an examination of the RGS data from XMM5, and is consistent with the \chandra\ LETG results found by \cite{Marshall+2003}. 

A variety of physical models; partial covering, soft-Comptonisation and blurred reflection,  all provide similar statistical fits to the data. Each of the models suggests a different physical interpretation for the variability. In the partial covering model, Fig.~\ref{fig:pcfcorr} shows that the lower flux observed in XMM4 and XMM5 is explained by an increase in column density and covering fraction in the ionised absorber. No change in the intrinsic power law spectrum is required, and changes in the neutral absorber show no evident correlations. For the soft-Comptonisation model, both the hard and soft Comptonisation components must vary to produce the observed spectra. For the dimmer XMM4 and XMM5, the fraction of power emitted below r$_{\rm cor}$ in the hard Compton component (fpl), the index of the hard power law, and the opacity of the soft corona all decrease, while the temperature of the soft component increases. These parameters increase and decrease accordingly as the spectra rise in flux. Finally, both blurred reflection models attribute the variability to normalisation changes in the coronal emission. 

While the best fit partial covering model was able to explain the observed spectral shape, it is the poorest fit of the four spectral models. It also cannot easily explain the observed changes in the spectra, as seen in the PCA of Section~\ref{sect:pca}. At these high column densities, more variability in the absorption spectrum is present at lower energies than at higher energies, so the entire model changes shape between data sets. This is also contrary to the findings from the variability analysis of Section~\ref{sect:var}, in which the flatness of the hardness ratios between epochs are more suggestive of a single spectral feature changing in normalisation. 

Additionally, from the fact that $\Delta \alpha_{\rm ox}$ values agreed with 0, it was hypothesized that the X-ray spectra would not be significantly absorbed. In the best-fitting partial covering model, however, the absorbers dim the source by a factor of $\sim5-10$. This significant amount of absorption in an apparently X-ray normal AGN is difficult to explain. This model also requires a complex, non-spherical symmetry of the absorbers to explain the lack of $6.4\kev$ emission expected from the measured absorption. 

A soft excess produced by soft-Comptonisation in a secondary, warm corona was proposed in previous works (\citealt{Marshall+2003,Guainazzi2004}). However, similarly to the partial covering interpretation, the variations in this model fail to reproduce the observed PCA. This is likely due to the fact that the soft-Comptonisation and hard Comptonisation components are almost independently responsible for the variability at low and high energies, which is contrary to the changes in normalisation of a single component suggested by the variability analysis and shape of the PCA. As seen in Section~\ref{sect:optx}, the model is also unable to explain the high fluxes observed in the UV band, even when testing different black hole spins. The $\Delta \alpha_{\rm ox}$ values suggest an X-ray normal state for Mrk 478 at all epochs, so it is unlikely that the UV data is intrinsically extreme.

One crucial disagreement between models is on the shape of the primary continuum. In the partial covering model, the power law slope is extremely steep, with $\Gamma = 2.99\pm0.02$ for all epochs. Although NLS1 galaxies typically feature steeper spectra than other AGN (e.g. \citealt{Boller+1996,Brandt+1997,Grupe+2001}), this value is high. In the soft-Comptonisation model, the spectrum is much flatter, with the index ranging between $2.0-2.2$ between epochs. These values are average among NLS1 AGN. The blurred reflection models require intermediate slopes, steeper than those of typical NLS1 galaxies, but not so extreme as the partial covering model. {\sc reflionx} suggests slopes in the $2.6-2.8$ range, and {\sc relxill} measures this to be lower, with slopes around $2.4-2.6$.

Although the variations in slopes between models are all able to explain the $0.3-10\kev$ spectra, they differ at $E > 10\kev$. The partial covering and blurred reflection models predict $15-100\kev$ fluxes below the BAT survey limit. Owing to the very shallow power law measured in the {\sc optxagnf} model, the predicted flux for this model is the highest ($\sim2\times10^{-12}$ \ergpscmps) and just at the threshold limit of the BAT survey (\citealt{Oh+2018}). All spectra are consistent with null detections using \suzaku\ PIN. 

One of the most curious features revealed by the spectral analysis is the unique iron line profile of Mrk 478. AGN typically show evidence of both narrow and broad $6.4\kev$ lines, with the broad line likely produced by reflection of the primary emission off of neutral material (e.g. the torus). However, no evidence for such a feature is detected in any of the presented models. There are some plausible explanations for this - the spectrum is very steep, meaning that a narrow line may be buried below the continuum at these high energies if the reflection off of neutral material was sufficiently weak. Additionally, only XMM5 has high signal-to-noise at these high energies; the other data sets suffer from short exposure and high background making it difficult to model narrow features at high energies.

The $\sim6.7\kev$ emission line, corresponding to \fexxv\ emission, is required by all models. For 2 additional free parameters, the fit statistic improves by $\Delta C=9$ for partial covering, up to $\Delta C=36$ using {\sc relxill}. Examining the residuals for all data before adding the line consistently shows an excess at this energy. The line energy is well constrained in all models, and is always in agreement with $6.7\kev$. An examination of the background of each spectrum does not suggest that the feature arises from improper background subtraction. The $6.7\kev$ emission line is narrow, and therefore likely does not originate in the inner accretion disc, as no evidence for relativistic broadening is seen. Instead, the feature may be the result of fluorescent emission lines from ionised layers in the inner region of the torus (\citealt{Matt+1996,Bianchi+2002,Costantini+2010}). It may also be produced in the broad or narrow line regions. 

Typically, the detection of $6.7\kev$ \fexxv\ emission is accompanied by an \fexxvi\ emission line at $6.97\kev$. However, no evidence for this feature is seen in the partial covering model. For the other models, the data are consistent with an emission feature at $6.97\kev$. Attempting to add the emission line to each data set only results in a significant detection for the dim flux, high quality XMM5 data. For this data set, the feature is particularly prominent when modelling the data with {\sc relxill}, where adding a narrow $6.97\kev$ feature improves the fit by $\Delta C = 12$ for one additional free parameter. However, additional higher quality data is required to confirm the detection of \fexxvi\ emission. If present, the emission likely originates from the same region as the \fexxv\ emission. 

%%%%%%%%%%%%%%%%%%%%%%%%%%%%%%%%%%%%%%%%%%%%%%%%%%%%%%%%%%%%%%%%%%%%%%%%%%%%%%%%%%%%%%%%%%%%%%%%%

%%%%%%%%%%%%%%%%%%%%%%%%%%%%%%%%%%%%%%%%%%%%%%%%%%%%%%%%%%%%%%%%%%%%%%%%%%%%%%%%%%%%%%%%%%%%%%%%%

%%%%%%%%%%%%%%%%%%%%%%%%%%%%%%%%%%%%%%%%%%%%%%%%%%%%%%%%%%%%%%%%%%%%%%%%%%%%%%%%%%%%%%%%%%%%%%%%%
%%%%%%%%%%%%%%%%%%%%%%%%%%%%%%%%%%%%%%%%%%%%%%%%%%%%%%%%%%%%%%%%%%%%%%%%%%%%%%%%%%%%%%%%%%%%%%%%%

\subsection{The Blurred Reflection Model}
\label{sect:discussionref}

Based on the low variability in hardness ratio throughout and between all epochs as well as the predicted and observed shapes of the PCA, the reflection model is the most likely physical explanation for the X-ray emission of Mrk 478. This reflection component was modelled using both  {\sc relxill} and {\sc reflionx} convolved with the blurring model {\sc kerrconv}. The results from these spectral fits differ in interpretation. {\sc relxill} predicts a highly ionised spectrum dominated by reflection off the inner accretion disc, while {\sc reflionx} suggests a low ionisation and power law dominated spectrum. However, based on the calculated $\alpha_{\rm ox}$ values (Section~\ref{sect:var}), Mrk 478 is behaving like a normal AGN and is not expected to display extreme spectral properties, like high values of R. In this sense, the {\sc reflionx} interpretation seems more consistent. 

The different measurements of ionisation and reflection fraction may be driven by intrinsic differences in the models themselves, discussed in detail in \cite{Garcia+2013}. In the reflection model, the soft excess is produced by a multitude of narrow emission lines which are then relativistically blurred. Between the two models, however, there are significant differences in the abundances of the elements responsible for these narrow emission lines, including O, Ne and Fe. This results in significant differences in the shape of the broad \feka\ line and, more notably, the soft excess. \cite{Garcia+2013} also note significant deviations between models at high ionisations, which although hard to explain, may also be attributable to the different abundances as well as different ionisation states of each element included in each model. 

Contrary to the differences in ionisation and reflection fraction, the measured values for spin and inclination between reflection models are comparable at the $90$ per cent level. The uncertainties on these values are only measurement errors.  \cite{Bonson+2016} perform extensive simulations based on {\sc relxill} parameters, and test the reproducibility of each model parameter. Although they consider a narrower energy range of $2.5-10\kev$, they find that the inclination can be reliably constrained to within $10\degree$, and spin to within $\sim10$ per cent above $a=0.9$. If these systematic error bars are adopted, spin and inclination values are in close agreement. Therefore, both reflection models suggest a maximum spinning black hole viewed at a low inclination.

The preference of the blurred reflection model to explain the observed spectrum of Mrk 478 also requires an explanation for the under-abundance of iron found by both {\sc reflionx} and {\sc relxill}. The discrepancy in best fit abundance between models can be explained by the different element abundances used in the two models. The iron abundance used in {\sc relxill} is $\sim30$ per cent lower than that of {\sc reflionx} (\citealt{Garcia+2013}), and when this is taken into account, the values agree within error. The low iron abundance is also in agreement with the results of \cite{Zoghbi+2008}. 

While many AGN require super-solar abundances to explain the reflection spectrum, very few require sub-solar abundances. For HE 0436--4717, \cite{Bonson+2015} find an iron abundance of $\sim0.4$ times solar value, and list possible causes such as low star formation rates yielding more pristine and metal-poor gas or a lack of type Ia supernovae (e.g. \citealt{Groves+2006}). Another interpretation presented by \cite{Skibo+1997} is cosmic ray spallation, wherein cosmic rays strike iron nuclei and cause it to lose nucleons, resulting in the creation of lower mass elements including Ti, V, Mn and Cr (\citealt{Skibo+1997,Gallo+2019spal}). \cite{Turner+2010} also propose this explanation for NGC 4051, another source that can be identified as having low iron (e.g. \citealt{Patrick+2012}).

%%%%%%%%%%%%%%%%%%%%%%%%%%%%%%%%%%%%%%%%%%%%%%%%%%%%%%%%%%%%%%%%%%%%%%%%%%%%%%%%%%%%%%%%%%%%%%%%%
%%%%%%%%%%%%%%%%%%%%%%%%%%%%%%%%%%%%%%%%%%%%%%%%%%%%%%%%%%%%%%%%%%%%%%%%%%%%%%%%%%%%%%%%%%%%%%%%%
%%%%%%%%%%%%%%%%%%%%%%%%%%%%%%%%%%%%%%%%%%%%%%%%%%%%%%%%%%%%%%%%%%%%%%%%%%%%%%%%%%%%%%%%%%%%%%%%%

\section{Conclusion}
\label{sect:conclusion}
A variability and spectral analysis for the NLS1 Mrk 478 is presented, using all available data from \xmm\ and \suzaku\ spanning from 2001 to 2017. These spectra differ in $0.3-10\kev$ flux by a factor of $\sim2$, while spectral shape does not appear to vary between observations. Data are well fitted by a variety of physical models; partial covering including one ionised and one neutral absorber, soft-Comptonisation using {\sc optxagnf}, and blurred reflection using both {\sc relxill} and {\sc reflionx} reflection models. However, through an analysis of the variability between data sets, it is revealed that the blurred reflection model best explains changes between epochs. This is especially apparent using PCA, when only the blurred reflection models can reproduce the flat shape of the first principal component. 

Although the two blurred reflection models disagree in some measurements, they both predict a rapidly rotating black hole seen at a shallow viewing angle. Both reflection models also suggest an under-abundance of iron, with values of $\sim0.5$ times solar abundances. All models support the existence of a narrow $6.7\kev$ emission line, which is attributable to \fexxv\ emission. XMM5 also shows some evidence for a narrow \fexxvi\ emission line at $6.97\kev$ using both reflection models and soft-Comptonisation. 

Mrk 478 has not been detected by the \swift\ BAT instrument, and a $72\ks$ \suzaku\ PIN observation resulted in a null-detection. There is some disagreement in the $10-100\kev$ fluxes predicted by the presented models, with soft-Comptonisation suggesting a much flatter and brighter spectrum at high energies than partial covering or blurred reflection models. Hard X-ray observations of Mrk 478 with \nustar\ may help distinguish further between spectral models, as well as provide another look at the $6-7\kev$ iron profile.

More consideration to the blurred reflection interpretation, as well as the iron profile, will be given in a companion to this work, in which a more detailed timing and time-resolved spectroscopic analysis will be given to the long \xmm\ exposure (XMM5). X-ray spectra obtained with higher resolution instruments such as \xrism\ (\citealt{xrismins}) and \athena\ (\citealt{athenains}) will allow for more detailed mapping of the iron profile, and possibly allow for the a clearer detection of ionised iron emission, narrow $6.4\kev$ emission, and confirm an under-abundance of iron in this unique NLS1.

\section*{Acknowledgements}
The \xmm\ project is an ESA Science Mission with instruments and contributions directly funded by ESA Member States and the USA (NASA). This research has made use of data obtained from the \suzaku\ satellite, a collaborative mission between the space agencies of Japan (JAXA) and the USA (NASA). We thank the referee for their helpful comments and suggestions which improved the original manuscript. AZ was partially supported by NASA under award 80NSSC18K0377. SGHW and LCG acknowledge the support of the Natural Sciences and Engineering Research Council of Canada (NSERC).

%%%%%%%%%%%%%%%%%%%%%%%%%%%%%%%%%%%%%%%%%%%%%%%%%%%
%%%%%%%%%%%%%%%%%%%%% REFERENCES %%%%%%%%%%%%%%%%%%

\bibliographystyle{mnras}
\bibliography{refs}

% Don't change these lines
\bsp
\label{lastpage}
\end{document}